\documentclass[aps,prl,twocolumn,groupedaddress]{revtex4-1}

\usepackage{graphicx, bm, amsfonts, amssymb, amsmath, appendix}
\usepackage{rotating}
\usepackage[caption=false]{subfig}

\usepackage[colorlinks=true,citecolor=blue,linkcolor=blue]{hyperref}

\newcommand{\U}{\mathcal U}

\begin{document}

\title{Implicit Ligand Theory: Rigorous Binding Free Energies and Thermodynamic Expectations from Molecular Docking}

\author{David D. L. Minh}
\email[Electronic Address: ]{dm225@duke.edu}

\affiliation{Department of Chemistry, Duke University, Durham NC 27708 USA}

\date{\today}

\begin{abstract}
A rigorous formalism for estimating noncovalent binding free energies and thermodynamic expectations from calculations in which receptor configurations are sampled independently from the ligand is derived.  
Due to this separation, receptor configurations only need to be sampled once, facilitating the use of binding free energy calculations in virtual screening.  
Demonstrative calculations on a host-guest system yield good agreement with previous free energy calculations and isothermal titration calorimetry measurements.
Implicit ligand theory provides guidance on how to improve existing molecular docking algorithms and insight into the concepts of induced fit and conformational selection in noncovalent macromolecular recognition.
\end{abstract}

\maketitle

\section{Introduction}

The goal of molecular docking is to predict the most stable configuration of a noncovalent complex between a ligand and receptor.  Based on this configuration, the complex is assigned a score which may be used to approximately rank the binding affinity of one ligand to the receptor versus another.  Molecular docking has many potential applications, and has been most prominently applied to the virtual screening \cite{Shoichet:2004p6935, Klebe:2006p6699} of chemical libraries to aid the development of pharmaceuticals.

Given the three-dimensional structure of a protein receptor, docking algorithms have proven reasonably adept at sampling stable conformations of small organic ligands in the complex.  Unfortunately, current scoring functions perform poorly at predicting binding free energies \cite{Kim:2008p6686, Moitessier:2009p6716, Plewczynski:2010p6700}.  Hence, docking is typically used to filter a large library of potential ligands to a smaller binder-enriched library that may be pursued experimentally or by more accurate and expensive computational methods \cite{Wang:2001p7494,Lin:2003p6839,Graves:2008p6534,Thompson:2008p7384,Hou:2010p6692}.  Even in this capacity, however, scoring functions are inconsistent, frequently presenting false positives (ligands predicted to bind but actually have weak or no affinity) and false negatives (ligands predicted not to bind but actually have significant affinity).  For example, docking programs often have difficulty distinguishing binding compounds from decoys in which the chemical connectivity has been randomized \cite{Kim:2008p6686,Chang:2010p7063}.  Improved scoring functions would increase the capability to discern binders from non-binders.

The improvement of scoring functions, however, has been hindered by the lack of a rigorous formalism for obtaining binding free energies from molecular docking.
While molecular docking calculations are usually performed with a rigid receptor, existing formalisms for binding free energies require a flexible receptor.
Here, I derive a formalism, \emph{implicit ligand theory}, for estimating binding free energies and thermodynamic expectations based on docking ligands to rigid receptor structures.
I also describe practical aspects of statistical estimation, present example calculations, and discuss how physics-based (opposed to empirical or knowledge-based) docking algorithms (see \cite{Huang:2006p6690}) may be modified to exploit it.
Beyond molecular docking, implicit ligand theory provides insight into the concepts of induced fit and conformational selection in noncovalent macromolecular recognition.

\section{Theory}

The standard binding free energy, the free energy of a noncovalent association between a receptor $R$ and ligand $L$ to form a complex $RL$, 
$R + L \rightleftharpoons RL$, is,
\begin{eqnarray}
\Delta G^\circ = -\beta^{-1} \ln \left( \frac{C^\circ C_{RL}}{C_R C_L} \right),
\end{eqnarray}
where $\beta = (k_B T)^{-1}$ is the inverse of Boltzmann's constant, $k_B$, times the temperature in Kelvin, $T$, $C^\circ$ is the standard concentration (typically 1 M), and $C_X$ is the equilibrium concentration of species $X \in \{R, L, RL\}$ \footnote{Activities have been assumed to be unity, a reasonable approximation in the limit of low concentrations.}.

Statistical thermodynamics relates the standard binding free energy to a ratio of configurational partition functions \cite{Gilson:1997p6409},
\begin{eqnarray}
\Delta G^\circ &=& -\beta^{-1} \ln \left( \frac{ Z_{RL,N} Z_N }{ Z_{R,N} Z_{L,N}} \frac{ C^\circ }{ 8 \pi^2 } \right) \label{eq:bindingFE_fullCI} \\
Z_{RL,N} &=& \int I_\xi e^{-\beta U(r_{RL},r_S)} dr_{RL} dr_S \label{eq:complexCI} \\
Z_{Y,N} &=& \int e^{-\beta U(r_Y,r_S)} dr_Y dr_S \label{eq:fullCI} \\
Z_N &=& \int e^{-\beta U(r_S)} dr_S,
\end{eqnarray}
in which symmetry numbers and a small pressure-volume term have been omitted from Eq. (\ref{eq:bindingFE_fullCI}).
$Z_{RL,N}$ and $Z_{Y,N}$ are configurational partition functions of the complex and of the species $Y \in \{R, L\}$, respectively, in $N$ molecules of solvent.
The potential energy $U(r_X,r_S)$ depends on $r_X$, the \emph{internal} coordinates of the receptor, ligand, or both in complex (the external degrees of freedom have been analytically integrated), and $r_S$, the coordinates of $N$ molecules of solvent.
The complex coordinates $r_{RL}$ may be decomposed into the internal coordinates of the receptor, $r_R$, and of the ligand, $r_L$, and six degrees of freedom describing their relative translation and rotation, $\xi_L$.
For simplicity, Jacobians for the transformation from Cartesian coordinates to a system with separated internal and external degrees of freedom are not shown in Eqs. (\ref{eq:complexCI}) and (\ref{eq:fullCI}).
In $Z_{RL,N}$, the indicator function $I_\xi \equiv I(\xi_L)$ takes values between 0 and 1 and determines whether the receptor and ligand are complexed or not.
For tight-binding complexes, the binding free energy is insensitive to the precise definition of $I_\xi$ \cite{Gilson:1997p6409}.

\subsection{Implicit Solvent Theory}

The configurational integrals in Eq. (\ref{eq:bindingFE_fullCI}) may be expressed in a formally equivalent but simpler form using implicit solvent theory \cite{Gilson:1997p6409}.  In implicit solvent theory, the \emph{interaction energy} is defined as $\psi(r_X,r_S) = U(r_X,r_S) - U(r_X) - U(r_S)$, where $U(r_X)$ is the potential energy of species $X$ by itself and $U(r_S)$ the potential energy of the solvent by itself.  By integrating the configurational partition functions over $r_S$, we may define the ratios,
\begin{eqnarray}
Z_{RL} & \equiv & \frac{Z_{RL,N}}{Z_N} = \int  I_\xi e^{-\beta [U(r_{RL}) + W(r_{RL})]} dr_{RL} 
\label{eq:Z_RL}\\
Z_Y & \equiv & \frac{Z_{Y,N}}{Z_N} = \int e^{-\beta [U(r_Y) + W(r_Y)]} dr_Y,
\label{eq:Z_Y}
\end{eqnarray}
where,
\begin{eqnarray}
W(r_X) = -\beta^{-1} \ln \left( 
\frac{ \int e^{-\beta \psi(r_X,r_S)} e^{-\beta U(r_{S}) } ~ dr_S}
       { \int e^{-\beta U(r_{S}) } ~ dr_S}
\right) \label{eq:solventPMF},~
\end{eqnarray}
is a potential of mean force that can be interpreted as the constant-pressure reversible work of transferring the species $X$ from the gas phase into the solvent.  In biomolecular modeling, $W(r_X)$ is frequently estimated as the sum of an electrostatic term from the Poisson-Boltzmann equation \cite{Wang:2008p6628} (or the Generalized Born approximation \cite{Feig:2004p6763}), and a non-electrostatic term, which to a first approximation is proportional to the molecular surface area.

In terms of implicit solvent configurational integrals, the standard binding free energy is,\begin{eqnarray}
\Delta G^\circ &=& -\beta^{-1} \ln \left( \frac {Z_{RL}}{Z_R Z_L} \frac{C^\circ}{8 \pi^2} \right).
\label{eq:FE_implicit}
\end{eqnarray}
As most implicit solvent models fail to account for specific interactions, such as hydrogen bonding, that can have important structural and energetic consequences, binding free energy calculations in implicit solvent are generally expected to be less accurate than those in explicit solvent \cite{Michel2008}.  Nevertheless, binding free energy calculations in implicit solvent have yielded promising agreement with experimental results (e.g. \cite{Chang2003,Chang2003a,Lee2005,Gallicchio2010}).

\subsection{Implicit Ligand Theory}

The development of implicit ligand theory is very similar to that of implicit solvent theory.
It involves defining the \emph{effective potential} as $\U(r_X) = U(r_X) + W(r_X)$, the \emph{effective interaction energy} as $\Psi(r_{RL}) = \U(r_{RL}) - \U(r_R) - \U(r_L)$, and,
\begin{eqnarray}
B(r_R) & = & -\beta^{-1} \ln \left( 
\frac{ \int  I_\xi e^{-\beta \Psi(r_{RL})} e^{-\beta \U(r_{L}) } ~ dr_L d\xi_L}
       { \int  I_\xi e^{-\beta \U(r_{L}) } ~ dr_L d\xi_L } \right) \nonumber \\
       &\equiv& -\beta^{-1} \ln \left< e^{-\beta \Psi} \right>^{r_L,\xi_L}_{L,I},
\label{eq:bindingPMF}
\end{eqnarray}
which is a potential of mean force that will subsequently be referred to as the \emph{binding PMF}.  Throughout this paper, angled brackets $\left< ... \right>^{r}_{X,...}$ will be used to denote an ensemble average over the coordinates $r$ listed in the superscript with respect to the density proportional to $q_{X,...}$, where $X$ describes the coordinates in the effective potential $\U(r_X)$, and $...$ are labels.
Here, $q_{L,I}(r_L,\xi_L) =  I_\xi e^{-\beta \U(r_L)}$.
Within angled brackets, I will use a shorthand notation in which functions implicitly depend on coordinates, e.g. $\Psi \equiv \Psi(r_{RL})$.

In terms of the binding PMF, Eq. (\ref{eq:FE_implicit}) may be written as,\begin{eqnarray}
\Delta G^\circ &=& 
-\beta^{-1} \ln \left( \frac{\int  I_\xi e^{-\beta \U(r_{RL})} dr_{RL} }
{\int e^{-\beta \U(r_{R})} dr_R \int e^{-\beta \U(r_L)} dr_L } \frac{C^\circ}{8 \pi^2} \right) \nonumber \\
&=& -\beta^{-1} \ln \left(  \frac{\int  I_\xi e^{-\beta [\U(r_R) + \Psi(r_{RL}) + \U(r_L)]} dr_{RL} }
{\int e^{-\beta \U(r_{R})} dr_R \int e^{-\beta \U(r_L)} dr_L } \frac{C^\circ}{8 \pi^2} \right) \nonumber \\
&=& -\beta^{-1} \ln \left( \frac{\int e^{-\beta [B(r_R)+  \U(r_{R})]} dr_R }
{\int e^{-\beta \U(r_{R})} dr_R } \frac{\Omega C^\circ}{8 \pi^2} \right) \nonumber \\
&\equiv& -\beta^{-1} \ln \left< e^{-\beta B} \right>^{r_R}_{R} + \Delta G_\xi\label{eq:FE_implicitL},
\end{eqnarray}
where $\Omega = \int  I_\xi d\xi_L$ (which may be analytically tractable) is the binding site volume, $\Delta G_\xi = - \beta^{-1} \ln \left( \frac{\Omega C^\circ}{8 \pi^2} \right)$ is the free energy of confining the ligand external degrees of freedom to the binding site, and $q_{R}(r_R) = e^{-\beta \U(r_R)}$.  Eqs. (\ref{eq:bindingPMF}) and (\ref{eq:FE_implicitL}) are the central theoretical results of this paper.

Implicit ligand theory provides a rigorous framework for binding free energies that separates the sampling of receptor and ligand configurations.
In Eq. (\ref{eq:FE_implicitL}), the receptor probability density is independent of any ligand configuration.
Likewise, the probability density of ligand internal coordinates in Eq. (\ref{eq:bindingPMF}) is independent from the receptor configuration.  
In practice, however, sampling from this ligand distribution may lead to slow convergence (this point will later be discussed in greater detail).
The primary benefit of implicit ligand theory is that the computationally expensive step of sampling receptor configurations only needs to be performed once.
Predicting binding free energies for a chemical library is then limited by the much faster process of sampling ligand conformations.

\subsection{Thermodynamic Expectations}

In addition to estimating the binding free energy, implicit ligand theory may also be used to estimate expected values of observables in the bound ensemble.  Observables may include, for example, the mean potential energy, interaction energy, or distance between a ligand and receptor atom.
Towards this end, it is useful to define a rigid-receptor expectation of an observable $O(r_{RL})$, weighted by the interaction energy,
\begin{eqnarray}
\Theta(r_R) & = &
\frac{ \int  I_\xi O(r_{RL}) e^{-\beta \Psi(r_{RL})} e^{-\beta \U(r_{L}) } ~ dr_L d\xi_L}
       { \int  I_\xi e^{-\beta \U(r_{L}) } ~ dr_L d\xi_L } \nonumber \\
& \equiv & \left< O e^{-\beta \Psi} \right>^{r_L,\xi_L}_{L,I}.
\label{eq:rrE}
\end{eqnarray}
If the observable is solely a function of the receptor configuration, then $\Theta(r_R)$ reduces to $O(r_R)e^{-\beta B(r_R)}$.

In terms of Eqs. (\ref{eq:bindingPMF}) and (\ref{eq:rrE}), the expectation of $O(r_{RL})$ with respect to the density proportional to $q_{RL,I}(r_{RL}) = I_\xi e^{-\beta \U(r_{RL})}$ is,
\begin{eqnarray}
\left< O \right>_{RL,I}^{r_{RL}} & \equiv &
\frac{ \int I_\xi O(r_{RL}) e^{-\beta \U(r_{RL})} dr_{RL} }
{ \int I_\xi e^{-\beta \U(r_{RL})} dr_{RL} } \nonumber \\
& = & 
\frac{ \int I_\xi O(r_{RL}) e^{-\beta [\U(r_R) + \Psi(r_{RL}) + \U(r_L)]} dr_{RL} }
{ \int I_\xi e^{-\beta [\U(r_R) + \Psi(r_{RL}) + \U(r_L)]} dr_{RL} } \nonumber \\
& = & \frac{ \int \Theta(r_R) e^{-\beta \U(r_R)} dr_R }{ \int e^{-\beta [B(r_R) + \U(r_R)]} dr_R } \nonumber \\
& \equiv & \frac{\left< \Theta \right>_R^{r_R}}{\left< e^{-\beta B} \right>_R^{r_R}} = \left< \Theta \right>_R^{r_R} e^{\beta [\Delta G^\circ - \Delta G_\xi]}
\label{eq:expectation}
\end{eqnarray}
Eqs. (\ref{eq:rrE}) and (\ref{eq:expectation}) significantly generalize \emph{implicit ligand sampling} \cite{Cohen2006}, a method to estimate the potential of mean force for the ligand center of mass.
The results of \citet{Cohen2006} may be obtained by choosing the observable as a Dirac delta function for the ligand center of mass, taking a natural logarithm, and multiplying by $\beta^{-1}$.
\citet{Cohen2006} applied implicit ligand sampling to study gas migration pathways in myoglobin, but the possibility of estimating other observables and binding free energies has not been previously recognized.

\section{Estimation}

Applying implicit ligand theory to predicting binding free energies involves three steps:
\begin{enumerate}
\item Sampling receptor configurations.
\item Estimating the binding PMF, $B(r_R)$, for each receptor configuration.
\item Estimating $\Delta G^\circ$ from $B(r_R)$ estimates.
\end{enumerate}
In this section, I present several ways, roughly in order of increasing complexity, that these steps may be accomplished.  A variant of one approach will be demonstrated later in the paper.

\subsection{Receptor Configurations}

Receptor configurations can be drawn from $q_R(r_R)$, any (possibly unnormalized) distribution $q_{R,w}(r_R)$ on the same support as $q_{R}(r_R)$ and for which $w(r_R) = q_{R}(r_R)/q_{R,w}(r_R)$ may be calculated, or from multiple distributions satisfying these conditions.  
Regardless of the sampling method, however, convergence of free energy estimates requires representative sampling of both the bound and unbound receptor configuration space.
A particularly straightforward protocol is to sample from the distribution proportional to $q_R(r_R) = e^{-\beta \U(r_R)}$; one conducts a molecular dynamics (MD) simulation in the implicit solvent used for $W(r_R)$, collecting snapshots at evenly spaced intervals that are longer than the statistical correlation time.  
This protocol may be satisfactory if receptor fluctuations are minimal and the ligand does not significantly perturb the receptor configurational ensemble.

For a receptor that undergoes larger structural fluctuations, sampling from multiple energetic minima may be facilitated by applying an external biasing potential (e.g. a harmonic bias) on one or more order parameters.  If it is known that a ligand significantly perturbs the receptor configurational ensemble, it can be useful to introduce multiple \emph{alchemical} intermediates into a simulation.
Alchemical calculations may involve a coupling parameter $\lambda$, defined such that the two groups (e.g. the receptor and ligand) are non-interacting at $\lambda=0$ and fully interacting with $\lambda=1$.  
Simulations are conducted with $\lambda$ at these end points and at multiple values in between.
Sampling in each stage may be enhanced by Hamiltonian replica exchange (e.g. \citet{Jiang2009,Gallicchio2010,Gallicchio2012}), which entails stochastically swapping the coordinates of different simulations with a probability that preserves the Boltzmann distribution.
Receptor configurations obtained through a flexible-receptor Hamiltonian replica exchange with a single ligand may subsequently be used for implicit ligand free energy calculations with other ligands in the chemical library.

As a caveat, implicit ligand theory does not provide a formal justification for docking to multiple experimentally determined structures (e.g. \cite{Rao:2008p6674}) or any other set of structures in which $w(r_R)$ is unknown (e.g. homology modeling or flexible docking).  One potential way to use information about multiple structures is to conduct multiple MD simulations with external potentials biased towards one or more of the structures.  To facilitate later analysis, the external potentials should be set up to promote overlap in the configuration space of different simulations.

\subsection{Estimating a Binding PMF}

The binding PMF $B(r_R)$ may be expressed in terms of a ratio of partition functions,
\begin{eqnarray}
B(r_R) &=& -\beta^{-1} \ln \left( 
\frac{ \int  I_\xi e^{-\beta \U(r_{RL})} ~ dr_L d\xi_L}
       { \int  I_\xi e^{-\beta [\U(r_{L}) + \U(r_R)] } ~ dr_L d\xi_L } \right),
\label{eq:bindingPMF_ratio}
\end{eqnarray}
which clarifies that $B(r_R)$ is a special type of free energy difference in which the receptor configuration $r_R$ is rigid.  Thus, $B(r_R)$ may be calculated using any one of many available methods to estimate free energy differences \cite{Chipot2007}, including free energy perturbation (FEP) \cite{Zwanzig1954}, thermodynamic integration (TI) \cite{Kirkwood1935}, and the Bennett Acceptance Ratio (BAR) \cite{BENNETT1976}.
While formally equivalent, free energy methods can have dramatically different convergence properties.

Based on the form of Eq. (\ref{eq:bindingPMF}), the most straightforward estimation protocol is FEP.  One can, for example, draw ligand configurations from the distribution proportional to $q_{L}(r_L,\xi_L) =  e^{-\beta \U(r_L)}$ by conducting a MD simulation of the ligand in the appropriate implicit solvent and collecting snapshots at sufficiently long intervals.
Because $q_L$ is independent of $\xi_L$, the external degrees of freedom sampled from the simulation may be replaced by a new $\xi_L$ sampled from the distribution proportional to $q_{\xi,I} =  I_\xi$.
The expectation in Eq. (\ref{eq:FE_implicitL}) may then be estimated by the sample mean,
\begin{eqnarray}
\hat{B}(r_R) = -\beta^{-1} \ln \frac{1}{N}
\sum_{n=1}^N e^{-\beta \Psi(r_{RL,n})},
\label{eq:bindingPMF_SM}
\end{eqnarray}
where $r_{RL,n}$ is the n$^{th}$ of N samples of the complex.  Throughout this paper, $\hat{A}$ will denote a statistical estimator  - an equation used to calculate a quantity based on sampled data.

In exponential averages such as Eq. (\ref{eq:bindingPMF_SM}), a small subset of samples may contribute a large portion of the sum.  The limiting case of an individual important sample inspires the severe \emph{dominant state approximation}, in which a single value of $\Psi(r_{RL})$ is used to estimate $B(r_R)$.
Exponential averages may also be estimated via a cumulant expansion \cite{Zwanzig:1954p6113}, here shown for Eq. (\ref{eq:bindingPMF}) to the fourth order,
\begin{eqnarray}
B(r_R) &\approx& 
\left< \Psi \right>_{L,I}^{r_L,\xi_L} 
- \frac{\beta}{2!} \left< \xi \Psi^2 \right>_{L,I}^{r_L,\xi_L}
+ \frac{\beta^2}{3!} \left< \xi \Psi^3 \right>_{L,I}^{r_L,\xi_L} \nonumber \\
&&- \frac{\beta^3}{4!} \left[\left< \xi \Psi^4 \right>_{L,I}^{r_L,\xi_L}-3 \left(\left< \xi \Psi^2 \right>_{L,I}^{r_L,\xi_L} \right)^2 \right],
\label{eq:1trajFEPexpansion}
\end{eqnarray}
where $\xi \Psi = \Psi(r_{RL}) - \left< \Psi \right>_{L,I}^{r_L,\xi_L}$.  Each expectation in the cumulant expansion may be estimated by the sample mean.

While formally correct, this approach to ligand sampling can converge slowly if most ligand configurations placed in the binding site have overlapping atoms and high values of $\Psi(r_{RL})$.  One potential solution to this problem is to sample the external degrees of freedom from a distribution biased towards energetically favorable orientations by a confining potential $U_c(\xi_L)$.  Multiplying and dividing Eq. (\ref{eq:bindingPMF}) by $\Omega_c = \int  I_\xi e^{-\beta U_c(\xi_L)} d\xi_L$ and the integrand in the numerator by $e^{-\beta U_c(\xi_L)}$ leads to,
\begin{eqnarray}
B(r_R) &=& -\beta^{-1} \ln \left< e^{-\beta [\Psi - U_c]} \right>^{r_L,\xi_L}_{L,Ic} - \beta^{-1} \ln \left( \frac{ \Omega_c }{\Omega} \right)
\label{eq:confinedBindingPMF}
\end{eqnarray}
where $q_{L,Ic} =  I_\xi e^{-\beta [\U(r_{L}) + U_c(\xi_L)] }$.  Good choices for $U_c(\xi_L)$, which may be ascertained from existing molecular docking algorithms (as will be discussed later in the paper), will favor the sampling of poses with low $\Psi(r_{RL})$.

Alternatively, the binding PMF may be calculated using the inverse form of Eq. \ref{eq:bindingPMF},
\begin{eqnarray}
B(r_R) &=& \beta^{-1} \ln \left( 
\frac{ \int  I_\xi e^{\beta \Psi(r_{RL})} e^{-\beta \U(r_{RL}) } ~ dr_L d\xi_L }
	{ \int  I_\xi e^{-\beta \U(r_{RL}) } ~ dr_L d\xi_L}
        \right) \nonumber \\
        &=& \beta^{-1} \ln \left< e^{\beta \Psi} \right>^{r_L,\xi_L}_{RL,I}.
\label{eq:IbindingPMF}
\end{eqnarray}
Ligand configurations from the distribution proportional to $q_{RL,I}(r_L,\xi_L) =  I_\xi e^{-\beta \U(r_{RL})}$ may be sampled, for example, from an implicit-solvent MD simulation in which the receptor is held rigid and the ligand is allowed to move, and the expectation estimated using the sample mean estimator.

This straightforward procedure is also problematic because of the rarity of sampling configurations in which the ligand is separated from the receptor or in which they overlap.  While these configurations are insignificant in the conformational ensemble in which receptor and ligand are fully interacting, they are relevant to the ensemble of noninteracting ligand and receptor, and the convergence of free energy differences requires phase space overlap between adjacent thermodynamic states \cite{Chipot2007}.  The phase space overlap problem may also be alleviated by calculating the free energy difference with a reference state in which the external degrees of freedom are confined,
\begin{eqnarray}
B(r_R) &=& \beta^{-1} \ln \left< e^{ \beta [\Psi - U_c]} \right>^{r_L,\xi_L}_{RL,I} 
       - \beta^{-1} \ln \frac{ \Omega_c }{\Omega}.
\label{eq:IconfinedBindingPMF}
\end{eqnarray}
The binding PMF may be estimated from the same samples as with Eq. (\ref{eq:IbindingPMF}), and will be more accurate the more closely $e^{-\beta U_c(\xi_L)}$ resembles the distribution of $\xi_L$ in the complex.

As discussed, phase space overlap problems are often resolved by introducing multiple alchemical stages into a calculation, and sampling may be enhanced by Hamiltonian replica exchange.  With multiple stages, the total free energy difference between states with $\lambda=0$ and $\lambda=1$ is the sum of free energy differences between adjacent stages, each of which may be estimated by FEP \cite{Zwanzig1954}, TI \cite{Kirkwood1935}, or BAR \cite{BENNETT1976}.  Alternatively, the total free energy difference may be estimated by the multistate Bennett Acceptance Ratio (MBAR) \cite{Shirts2008}.

\subsection{Estimating the Binding Free Energy}

Once $\hat{B}(r_R)$ is evaluated for each receptor configuration, the binding free energy may be calculated by estimating an ensemble average.  The appropriate method for estimating $\Delta G^\circ$ depends on how the receptor configurations $r_{R}$ are sampled.  If they are drawn from the distribution $q_R(r_R)$, then the expectation in Eq. (\ref{eq:FE_implicitL}) may be estimated by the sample mean,
\begin{eqnarray}
\Delta \hat{G}^\circ = -\beta^{-1} \ln \frac{1}{N} \sum_{n=1}^N e^{-\beta \hat{B}(r_{R,n})} + \Delta G_\xi,
\label{eq:imLigFE_SM}
\end{eqnarray}
in which $\hat{B}(r_{R,n})$ is the estimated binding PMF for the n$^{th}$ of N receptor configurations.  Because the implicit-ligand expression for the binding free energy, Eq. (\ref{eq:FE_implicitL}), has the same form as Eq. (\ref{eq:bindingPMF}), the dominant state approximation and cumulant expansion may also be applied.  

If receptor configurations are drawn from a biased distribution, the importance sampling identity,
\begin{eqnarray}
\left< O \right>_T &=& \frac{\int O(r) q_T(r) dr}{\int q_T(r) dr} \nonumber \\
&=& \frac{\int O(r) w(r) q_S(r) dr}{\int w(r) q_S(r) dr} = \frac{\left< w O \right>_S}{\left< w \right>_S},
\label{eq:Isample}
\end{eqnarray}
may be applied.  In this generic expression, $w(r) = q_T(r)/q_S(r)$ is a ratio of unnormalized densities $q_T(r)$ for the target distribution and $q_S(r)$ for the sampling distribution.
Using the sample mean estimator and importance sampling identity for the expectation in Eq. (\ref{eq:FE_implicitL}) leads to,
\begin{eqnarray}
\Delta \hat{G}^{\circ} = 
-\beta^{-1} \ln \frac{\sum_{n=1}^N w(r_{R,n}) e^{-\beta \hat{B}(r_{R,n})}}{\sum_{n=1}^N w(r_{R,n})} + \Delta G_\xi
\label{eq:imLigFE_Isample}
\end{eqnarray}
If receptor configurations are drawn from multiple biased distributions, then the expectation may be estimated using MBAR \cite{Shirts2008}.

\subsection{Thermodynamic Expectations}

Thermodynamic expectations may be estimated from the same data as the binding free energy.  
The appropriate estimator for $\Theta(r_R)$ will depend on how the ligand configurations were sampled.
Once $\Theta(r_R)$ is estimated for every sampled receptor configuration, the appropriate estimator for the expectation in Eq. (\ref{eq:expectation}) similarly depends on how the receptor configurations were sampled.
In the simplest case for $\Theta(r_R)$, if ligand configurations are sampled from $q_{\xi,I}$, then $\Theta(r_R)$ may be estimated by a sample mean.
In other cases, $\Theta(r_R)$ and $\left< O \right>^{r_{RL}}_{RL,I}$ may be estimated using importance sampling, MBAR \cite{Shirts2008}, or a combination thereof.

\section{Demonstration}

As a demonstration, implicit ligand theory calculations were performed to estimate the standard binding free energy of various ligands to Cucurbit[7]uril (CB[7]) in water.  
The binding of CB[7] to a number of ferrocenes, adamantanes, and bicyclooctanes has been well-characterized by both isothermal calorimetry and second-generation mining minima (M2) \cite{Chang2003,Chang2003a} free energy calculations \cite{Moghaddam2009,Moghaddam2011}.
Receptor configurations were sampled by molecular dynamics,
binding PMFs estimated with a multi-stage alchemical calculation and MBAR \cite{Shirts2008},
and the binding free energy calculated using Eq. (\ref{eq:imLigFE_SM}) or the dominant state approximation.

\subsection{Methods}

Molecular dynamics simulations at 300 K were performed with a slightly modified \footnote{Using the linear combination of pairwise overlap \cite{Weiser1999} algorithm, NAMD 2.9 calculates a negative surface area for CB[7].  NAMD directly uses Appendix B of \citet{Weiser1999}, in which the P$_1$ parameter for the N \emph{sp3} atom type with 1 bonded neighbor is $7.8602 \times 10^{-2}$, which is smaller than other P$_1$ values and the corresponding P$_2$ parameter.  By definition, P$_1$ should be larger than P$_2$.  To bring this parameter in line with other P$_1$ and to make it larger than P$_2$, this parameter was multiplied by 10.  The modified code yields a positive surface area for CB[7].} compilation of NAMD  \cite{Phillips2005} version 2.9.  When appropriate, CB[7] was fixed using the fixedAtoms parameter.  The ``commercial'' force field parameters and topologies from \citet{Moghaddam2011} were used for both CB[7] and its ligands.  To match the force field from \citet{Moghaddam2011} as closely as possible, 1-4 electrostatics were scaled by 0.5 and the nonbonded cutoff was set to 999 \AA, which effectively turns off cutoffs.  Water was represented with the Generalized Born Surface Area (GBSA) implicit solvent model without ions and a surface tension of 0.006 kcal/mol/\AA$^2$.  The receptor dielectric was 1.0 and solvent dielectric was 78.5. A time step of 1 fs (using a 2 fs time step with fixed atoms led to unstable trajectories) was used with Langevin dynamics.

CB[7] was minimized for 2500 steps and thermalized by increasing the temperature by 10 K and reinitializing velocities every 100 steps from 0 to 300 K.  Receptor snapshots were saved every 0.1 ns from a trajectory of 10 ns.  

Binding PMFs for every ligand in \citet{Moghaddam2011} with the minimized CB[7] structure (15 repetitions each) and 100 receptor simulation snapshots (1 repetition each) were estimated using Hamiltonian replica exchange, which can simultaneously dock a ligand and compute its binding free energy \cite{Gallicchio2010,Gallicchio2012}.
The implementation is similar to that from \citet{Gallicchio2012}, except that the receptor configuration is fixed.
A reservoir of ligand configurations \cite{Gallicchio2012} was generated by simulating the ligand for up to 10 ns and saving snapshots every 10 ps.
Simulations of the complex in which $\lambda$ controls the extent of interaction between the ligand and receptor were run with 
$\lambda \in \{0,10^{- 5},10^{-4},10^{-3},10^{-2},$
$0.1,0.2,0.3,0.4,0.5,$
$0.6,0.7,0.8,0.9,0.95,1.0\}$.
As implemented in NAMD, intermediate values of $\lambda$ used a soft-core potential with a van der Waals shift coefficient of 5.  Electrostatic interactions were turned on when $\lambda = 0.5$.  Using the colvars module, a flat-bottom harmonic potential with a spring constant of 10 kcal mol$^{-1}$ \AA$^{-1}$ and starting at 0.75 \AA~ was used to restrain the center-of-mass distance between the ligand core (heavy atoms except for the R groups in \citet{Moghaddam2011}) and the receptor heavy atoms.  This potential keeps the ligand within the binding site when interactions are turned off.
The binding site volume, $\Omega = \int  I_\xi d\xi_L$, is approximated as $4/3 \pi (0.75^3) (8\pi^2)$.
Because NAMD does not allow the simultaneous use of alchemical decoupling and implicit solvent, simulations were conducted in vacuum.

The replica exchange simulation was initiated by taking a random ligand configuration, applying a random rotation, and randomly placing it within the binding site.  This initial configuration was minimized  and thermalized with the same protocol as with CB[7], except that it was done in vacuum.  The thermalized structure was used to start each replica.  Occasionally, the random placement of the ligand led to high forces that caused the simulations to crash; in this case, the simulation was restarted with a different random initial configuration.

After every 5 ps of simulation for every value of $\lambda$, 1000 replica exchanges were attempted between each pair of adjacent $\lambda$ windows.  After each set of replica exchange attempts, the ligand configuration for $\lambda=0$ was replaced with a random ligand configuration from the reservoir, randomly rotated, and placed in the binding site.  (This type of reservoir swap satisfies detailed balance.)  The simulation was conducted for 25 cycles, saving snapshots every 0.5 ps, for a total of 2 ns of simulation for each binding PMF.
The docking and equilibration period, defined as the time before the potential energy of the fully coupled state is within 20 $k_B T$ of its energy for the final snapshot, was ignored in subsequent analysis.

Because alchemical coupling calculations were performed in vacuum, binding PMFs were estimated based on a decomposition of $B(r_R)$,
\begin{eqnarray}
B(r_R) &=& B_{cpl} + B_{RL} - B_L - \Delta U(r_R)
\label{eq:bindingPMF_decomposition}\\
B_{cpl} &=& -\beta^{-1} \ln \left( 
\frac{ \int  I_\xi e^{-\beta U(r_{RL})} dr_L d\xi_L }
       { \int  I_\xi e^{-\beta [U(r_L) + U(r_R)]} dr_L d\xi_L } \right) \nonumber \\
B_{RL} &=& -\beta^{-1} \ln \left( 
\frac{\int  I_\xi e^{-\beta \Delta U(r_{RL})} e^{-\beta U(r_{RL})} dr_L d\xi_L}
       { \int  I_\xi e^{-\beta U(r_{RL})} dr_L d\xi_L } \right) \nonumber \\
B_L &=& -\beta^{-1} \ln \left( 
\frac{ \int  I_\xi e^{-\beta \Delta U(r_L)} e^{-\beta U(r_L)} dr_L d\xi_L }
       { \int  I_\xi e^{-\beta U(r_L)} dr_L d\xi_L } \right). \nonumber
\end{eqnarray}
$B_{cpl}$ is the free energy of turning on the interactions between the ligand and the rigid receptor in vacuum.
$B_{RL}$, $B_L$, and $\Delta U(r_R)$ are free energies of transferring the complex, ligand, and receptor, respectively, from vacuum to the target state (in implicit solvent).
They are based on $\Delta U(r_X) = U_T(r_X) - U(r_X)$, the potential energy difference between $r_X$ in the target state versus the state from which configurations were sampled (in vacuum).
$B_{cpl}$ was estimated by applying MBAR \cite{Shirts2008} to snapshots from every 0.5 ps of simulation, and $B_{RL}$ and $B_L$ by single-step FEP (evaluating transfer free energies by MBAR would require calculating target-state potential energies for every snapshot using computationally expensive force fields).

This decomposition makes it straightforward to evaluate $B(r_R)$ for a variety of force fields using the same configurational samples.  
In this work, four are compared:
\begin{enumerate}
\item NAMD: the total potential energy from using GBSA in NAMD \cite{Phillips2005};
\item M2: the total potential energy from using the GBSA model in the M2 program \cite{Chang2003,Chang2003a};
\item PB: Poisson-Boltzmann electrostatic solvation free energies from UHBD \cite{Davis1991} and bond, angle, dihedral, coulomb, and van der Waals energies from the M2 program \cite{Chang2003,Chang2003a};
\item PBSA: Poisson-Boltzmann electrostatic solvation free energies from UHBD \cite{Davis1991} and bond, angle, dihedral, coulomb, van der Waals, and nonpolar surface area energies from M2 \cite{Chang2003,Chang2003a}, the combination used in \citet{Moghaddam2011}.
\end{enumerate}
During this step, the NAMD, M2, and UHBD programs are used strictly for single-point energy evalulations, not for minimization or dynamics.
Poisson-Boltzmann energies were calculated with a grid spacing of 0.18 \AA~with dimensions such that the maximum dimensions of the molecule are 0.7 (or less) of the final grid \cite{Moghaddam2011}.
For comparison, binding PMFs were also calculated from the dominant state approximation with PBSA energies, using the lowest value of $\Psi(r_{RL})$ observed in the simulations with $\lambda=0$ or $\lambda=1$.

Because receptor configurations were sampled from a simulation in GBSA implicit solvent, binding free energies were estimated by using Eq. (\ref{eq:imLigFE_Isample}).  Binding free energies were also estimated with the dominant state approximation: using the lowest observed value of $\hat{B}(r_R)$ to estimate $-\beta^{-1} \ln \left< e^{-\beta B} \right>^{r_R}_{R}$.

To demonstrate the calculation of thermodynamic expectations and for comparison with results from \citet{Moghaddam2011}, the mean values of six PBSA energies - van der Waals, coulomb, electrostatic solvation, valence (bond + angle + dihedral), nonpolar solvation, and total - were estimated for the complex, the receptor, and the ligand.  
Mean PBSA energies for the ligand and receptor were estimated by applying the importance sampling identity to the ligand from the non-interacting system in vacuum and to the receptor from the GBSA simulation, respectively. 
Occasionally, energies in the ligand trajectory briefly spiked to very high values.
In estimating the mean PBSA energies, these spikes were filtered out by removing data points in which the total PBSA energy is at least 100 $k_B T$ larger than the PBSA energy of the final snapshot.
As the spikes were likely caused by the finite molecular dynamics time step, they would probably be avoided by using a propagator the exactly preserves the Boltzmann distribution, e.g. Hybrid Monte Carlo \cite{DUANE1987}.

Towards estimating the mean PBSA energies of the complex, rigid-receptor expectations were estimated by applying MBAR \cite{Shirts2008} to snapshots from the non-interacting and fully interacting states,
\begin{eqnarray}
\hat{\Theta}(r_R) &=& \frac{ \sum_{n=1}^N w(r_{RL,n}) O(r_{RL,n}) }{ \sum_{n=1}^N w(r_{RL,n}) } \\
w(r_{RL}) &=& \frac{e^{-\beta (U_{PBSA}(R_L) - U_0(R_L))}}{1 + \frac{N_1}{N_0} e^{-\beta(U_1(r_{RL})-\bar{B}_{cpl} - U_0(r_{RL})} }, \nonumber
\label{eq:MBAR_expectation}
\end{eqnarray}
where $U_0(r_{RL})$ and $U_1(r_{RL})$ are the potential energies of the non-interacting and fully interacting complexes, respectively, $U_{PBSA}(r_L)$ is the PBSA energy of only the ligand, and $r_{RL,n}$ is the n$^{th}$ of $N$ snapshots of either the non-interacting ($N_0$ snapshots) or fully interacting complex ($N_1$ snapshots).  $\hat{B}_{cpl}$ was estimated by using MBAR \cite{Shirts2008} with all replicas.  While it would be possible to estimate the mean PBSA energies using all snapshots from all replicas, this was avoided because of the computational expense of Poisson-Boltzmann calculations, which can take over a minute per snapshot.  After obtaining $\hat{\Theta}(r_R)$, the importance sampling identity, Eq. (\ref{eq:Isample}), was used to estimate the expectations in Eq. (\ref{eq:expectation}).  To ensure consistency of the estimator - an estimate of a constant yields the same constant - $\hat{\Theta}(r_R)$ was calculated for $O = 1$, in which case $\left< \hat{\Theta}(r_R) \right>^{r_R}_R = \left<e^{-\beta B}\right>^{r_R}_R$.  This estimate of $\left<e^{-\beta B}\right>^{r_R}_R$ was used in the denominator of Eq. (\ref{eq:expectation}).

\subsection{Results}


Highlighting the importance of an accurate molecular mechanics model, binding PMF estimates are strongly dependent on the force field, as shown in Table \ref{tab:B}.  
For the large and highly charged bicyclooctane B11, switching the force field causes the binding PMF to change nearly 40 kcal/mol!
With increasing magnitude of charge, larger coulomb energies lead to larger values of $B_{cpl}$ and larger electrostatic solvation free energies increase the magnitude of $B_{RL}$, $B_L$, and $\Delta U(r_R)$ (for estimates of $B_{cpl}$, $B_{RL}$, and $B_L$, see Table \ref{tab:S1} of the Supplemental Material.
Thus, estimating the binding PMF with Eq. (\ref{eq:bindingPMF_decomposition}) entails the difficult task of computing a relatively small difference between large values.  
The importance of the force field has also been noted for M2 calculations \cite{Chang2003,Chang2003a}.
An alternate implementation, e.g. conducting replica exchange within implicit solvent rather than vacuum, may not require the implicit solvent model to be as accurate.

\begin{table*}[h]
\center
\begin{tabular}{l | c c c c | c}
Ligand & NAMD & M2 & PB & PBSA & min$\{\Psi(r_{RL})\}$  \\ 
\hline
AD1 & -14.1 (0.79) & -22.0 (0.51) & -23.0 (0.82) & -25.5 (0.83) & -31.3 (0.55) \\ 
AD2 & -32.5 (0.15) & -29.0 (0.13) & -26.8 (0.12) & -29.4 (0.12) & -36.9 (0.30) \\ 
AD3 & -31.0 (0.16) & -30.7 (0.18) & -28.9 (0.23) & -31.6 (0.23) & -40.3 (0.28) \\ 
AD4 & -44.0 (0.94) & -36.7 (1.11) & -24.0 (1.12) & -26.9 (1.12) & -36.1 (0.45) \\ 
AD5 & -32.2 (0.68) & -29.0 (0.25) & -26.0 (0.14) & -28.5 (0.14) & -36.2 (0.29) \\ 
B02 & -12.8 (0.41) & -18.8 (0.38) & -19.8 (0.53) & -22.6 (0.53) & -30.6 (0.54) \\ 
B05 & -40.4 (0.29) & -30.6 (0.40) & -19.5 (0.50) & -22.3 (0.50) & -34.3 (0.63) \\ 
B11 & -52.4 (1.50) & -38.5 (1.81) & -14.1 (1.72) & -17.5 (1.63) & -39.3 (1.90) \\ 
F01 & -1.8 (1.57) & -5.5 (0.81) & -10.9 (0.53) & -13.6 (0.53) & -24.7 (0.33) \\ 
F02 & -14.8 (0.96) & -14.1 (0.67) & -16.2 (0.42) & -19.2 (0.42) & -31.7 (0.38) \\ 
F03 & -16.3 (1.61) & -13.1 (1.03) & -16.4 (0.95) & -19.5 (0.94) & -31.1 (0.71) \\ 
F06 & -30.6 (0.18) & -20.0 (0.21) & -21.9 (0.18) & -25.4 (0.18) & -37.2 (0.24) \\ 
\hline
R$^2_{ITC}$  & 0.884 & 0.750 & 0.454 & 0.490 & 0.883 \\ 
RMSE$_{ITC}$ &  12.8 &   7.9 &   4.9 &   4.7 &  12.2 \\ 
\hline
R$^2_{Gilson}$  & 0.827 & 0.907 & 0.705 & 0.712 & 0.792 \\ 
RMSE$_{Gilson}$ &  10.4 &   4.8 &   5.4 &   4.5 &  11.3 \\ 
\end{tabular}
\caption{The mean and standard deviation of 15 independent estimates of the binding PMF, $B(r_R)$, (kcal/mol) for various ligands to the minimized structure of CB[7], based on applying Eq. (\ref{eq:bindingPMF_decomposition}) with different force fields (NAMD, M2, PB, and PBSA columns) or on using the minimum observed value of the interaction energy $\Psi(r_{RL})$ from PBSA energies during the $\lambda=0$ and $\lambda=1$ simulations (min$\{\Psi(r_{RL})\}$ column).
The bottom rows show the correlation coefficient (R$^2$) and root mean square error (RMSE, Eq. (\ref{eq:RMSE})) with respect to isothermal titration calorimetry experiments (ITC) and mining minima calculations (Gilson) from \citet{Moghaddam2011} that result from the dominant state approximation - calculating $\Delta \hat{G}^\circ$ by using a single binding PMF estimate $\hat{B}(r_R)$ as an estimate for $-\beta^{-1} \ln \left< e^{-\beta B} \right>_R^{r_R}$ in Eq. (\ref{eq:FE_implicitL}).
\label{tab:B}}
\end{table*}

With 2 ns of total simulation for all replicas, the standard deviation of binding PMF estimates ranges from 0.12 to 1.63 kcal/mol (Table \ref{tab:B}), with most estimates on the lower range of imprecision.  
For all of the components of Eq. (\ref{eq:bindingPMF_decomposition}), the mean estimate does not appear to shift after about 0.75 ns, and additional sampling reduces the standard deviation of the estimate (see Fig. \ref{fig:Bconvergence} and Fig. \ref{fig:S1} in the Supplemental Material.
There is no unique component that limits the convergence of $\hat{B}(r_R)$; the slowest converging component varies from ligand to ligand.
The binding PMF estimate $\hat{B}(r_{R})$ and the minimal interaction energy $\min\left\{\Psi(r_{RL})\right\}$ converge at about the same rate, suggesting that the limiting factor for convergence is finding a configuration with the lowest interaction energy.
This interpretation is corroborated by the fact that largest ligands with the most rotatable bonds (see \citet{Moghaddam2011} for structures) also have the most variance in $\hat{B}(r_R)$, as the flexibility increases the challenge of finding configurations with low $\Psi(r_{RL})$.

\begin{figure}
\includegraphics{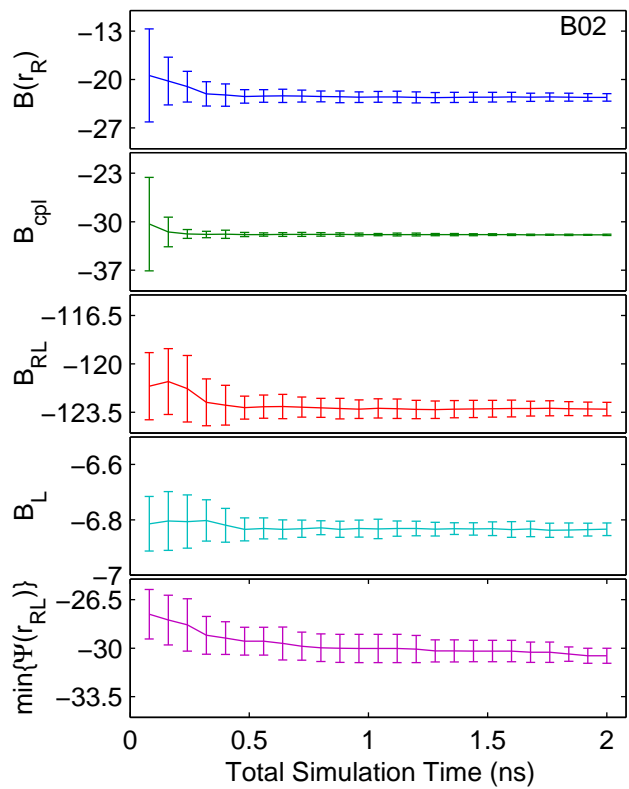}
\caption{The mean and standard deviation of 15 independent estimates of $B(r_R)$, $B_{cpl}$, $B_{RL}$, $B_L$, and min$\left\{\Psi(r_{RL})\right\}$ (kcal/mol) based on PBSA energies as a function of total MD simulation time for the ligand B02.  Analogous plots for the other ligands in this study are available as Fig. \ref{fig:S1} in the Supplemental Material. \label{fig:Bconvergence}}
\end{figure}


The accuracy of binding free energy estimates was assessed with the correlation coefficient and root mean square error,
\begin{eqnarray}
\textrm{RMSE}(m1,m2) = \sqrt{ \frac{1}{L} \sum_{l=1}^{L} (\Delta G^\circ_{l,m1} - \Delta G^\circ_{l,m2})^2 }
\label{eq:RMSE}
\end{eqnarray}
between methods $m1$ and $m2$, where $\Delta G^\circ_{l,m}$ is the binding free energy estimate for ligand $l$ of $L$ ligands using method $m$ (Tables \ref{tab:B} and \ref{tab:FE}).

Binding free energy estimates based on the binding PMF for a minimized receptor structure suffices to provide high correlation with experiment (R$^2 = 0.884$ for NAMD) and M2 free energy calculations (R$^2 = 0.827$ for NAMD) (see Table \ref{tab:B}).  Surprisingly, binding free energies from NAMD GBSA calculations are more highly correlated to these benchmarks than $\Delta \hat{G}^{\circ}$ from PBSA calculations.  Ironically, the high correlation may be explained by inaccurately large binding PMF values resulting from highly charged ligands, as the molecules in this set with the strongest charges also tend to have stronger binding affinities.
Although the correlation coefficient is high, the RMSE is also considerable, over 10 kcal/mol.
Similar performance (R$^2$ and RMSE) is observed by using the dominant state approximation with PBSA calculations.
In contrast, using Eq. (\ref{eq:bindingPMF_decomposition}) with PBSA leads to less correlated (lower R$^2$) but more accurate (lower RMSE) estimates of the binding free energy.

Even for this simple system, binding free energy estimates are substantially improved by using multiple receptor structures (Table \ref{tab:FE}).  With binding PMFs from PBSA energies for 100 receptor structures, there is both higher correlation and lower RMSE with respect to experiment (R$^2_{Exp} = 0.704$, RMSE$_{Exp}$ = 4.5) and especially with respect to M2 free energy calculations (R$^2_{Gilson} = 0.925$, RMSE$_{Gilson}$ = 2.4).    

\begin{table*}
\center
\begin{tabular}{l | c | c | c c c c }
Ligand & ITC & Gilson & NAMD & M2 & PB & PBSA \\ 
\hline
AD1 & -14.1 & -18.2 &  -9.4 (0.23) & -16.3 (0.15) & -17.6 (0.25) & -20.1 (0.25) \\ 
AD2 & -19.4 & -25.9 & -27.9 (0.19) & -24.3 (0.22) & -22.9 (0.27) & -25.4 (0.26) \\ 
AD3 & -20.4 & -25.6 & -35.7 (5.03) & -28.6 (1.87) & -23.5 (0.23) & -26.2 (0.23) \\ 
AD4 & -21.5 & -29.7 & -40.5 (0.21) & -33.7 (0.32) & -24.3 (1.11) & -27.1 (1.06) \\ 
AD5 & -19.1 & -24.1 & -29.5 (1.24) & -24.0 (0.20) & -22.0 (0.35) & -24.4 (0.34) \\ 
B02 & -13.4 & -12.0 &  -9.0 (0.38) & -13.7 (0.16) & -15.4 (0.26) & -18.1 (0.25) \\ 
B05 & -19.5 & -23.1 & -38.0 (0.40) & -27.7 (0.27) & -18.6 (0.27) & -21.4 (0.27) \\ 
B11 & -20.6 & -22.4 & -51.2 (0.34) & -37.3 (0.24) & -17.2 (0.53) & -20.5 (0.51) \\ 
F01 & -12.9 & -10.2 &   0.3 (0.82) &  -0.6 (0.34) &  -4.9 (0.26) &  -7.6 (0.25) \\ 
F02 & -16.8 & -12.4 & -12.0 (0.70) &  -9.6 (0.75) & -11.7 (0.70) & -14.6 (0.71) \\ 
F03 & -17.2 & -12.2 & -10.2 (0.16) &  -7.3 (0.24) & -10.2 (0.22) & -13.2 (0.22) \\ 
F06 & -21.0 & -17.8 & -24.1 (0.34) & -14.1 (0.46) & -16.2 (0.51) & -19.7 (0.52) \\ 
\hline
R$^2_{ITC}$ &  & 0.782 & 0.870 & 0.745 & 0.671 & 0.704 \\ 
RMSE$_{ITC}$ &  &   4.6 &  14.0 &   9.0 &   4.4 &   4.5 \\ 
\hline
R$^2_{Gilson}$ &  &  & 0.841 & 0.892 & 0.923 & 0.925 \\ 
RMSE$_{Gilson}$ &  &  &  11.3 &   5.9 &   3.4 &   2.4 \\ 
\end{tabular}
\caption{Estimates of the binding free energy $\Delta G^\circ$ (kcal/mol) of various ligands to CB[7].  First, binding PMFs $B(r_R)$ are estimated based on Eq. (\ref{eq:bindingPMF_decomposition}) for 100 receptor snapshots from a simulation in GBSA implicit solvent.  Then $\Delta \hat{G}^\circ$ is calculated using Eq. (\ref{eq:imLigFE_Isample}).
The value in the parentheses is the standard deviation from bootstrapping: the binding free energy is estimated based on 1000 random selections of 100 binding PMFs.
The experimental and Gilson columns are isothermal calorimetry measurements (ITC) and M2 calculations, respectively, taken from \citet{Moghaddam2011}.
The bottom rows are the correlation coefficient (R$^2$) and root mean square error (RMSE, Eq. (\ref{eq:RMSE})) with respect to the ITC and Gilson columns.
\label{tab:FE}}
\end{table*}

While there are some variations on the order of a few kcal/mol, mean potential energy changes upon complexation are also consistent with results from \citet{Moghaddam2011} (Table \ref{tab:PE}).
Minor discrepencies between M2 and implicit ligand free energy and mean potential energy calculations may be explained by a combination of imperfect sampling in the current calculations and the approximations in M2.
As the described calculations were performed in vacuum, the samples may not be from the same configurational space as those in implicit solvent.
On the other hand, M2 assumes that the energy landscape of the ligand, receptor, and complex are a truncated harmonic wells with anharmonicity corrections.

\begin{table*}
\center
\begin{tabular}{l | c c c c c c }
Ligand & VDW & Coul & PB & Val & NP & Total \\ 
\hline
AD1 & -32.5 (0.471) &   0.1 (1.509) &   4.8 (1.547) &  -5.0 (2.785) &  -2.5 (0.011) & -35.2 (2.547) \\ 
AD2 & -33.6 (0.931) & -65.8 (1.032) &  64.9 (0.783) &  -5.9 (1.910) &  -2.5 (0.017) & -42.9 (1.693) \\ 
AD3 & -32.8 (0.718) & -64.4 (0.693) &  62.2 (0.855) &  -5.7 (2.128) &  -2.6 (0.009) & -43.4 (2.388) \\ 
AD4 & -38.1 (1.400) & -125.2 (3.283) & 124.4 (1.003) &   1.9 (4.475) &  -2.7 (0.070) & -39.9 (2.817) \\ 
AD5 & -33.3 (1.374) & -65.1 (1.549) &  64.8 (1.415) &  -4.9 (1.782) &  -2.5 (0.025) & -40.9 (1.834) \\ 
B02 & -33.3 (0.622) &  -5.8 (1.067) &   9.7 (0.770) &   1.4 (3.379) &  -2.7 (0.022) & -30.6 (2.187) \\ 
B05 & -32.9 (0.896) & -138.2 (1.231) & 138.0 (1.151) &  -2.3 (1.236) &  -2.8 (0.013) & -38.1 (1.673) \\ 
B11 & -39.9 (1.192) & -199.3 (2.280) & 212.0 (1.066) &  -5.6 (5.431) &  -3.4 (0.075) & -36.2 (4.475) \\ 
F01 & -26.2 (0.497) &  -8.2 (1.824) &  14.2 (1.119) &   8.7 (4.842) &  -2.7 (0.017) & -14.3 (4.079) \\ 
F02 & -26.9 (1.518) & -65.7 (2.078) &  65.9 (0.923) &  -0.9 (2.237) &  -3.0 (0.012) & -30.6 (2.253) \\ 
F03 & -28.7 (0.832) & -58.0 (0.987) &  64.2 (0.649) &  -0.4 (3.552) &  -3.0 (0.015) & -26.1 (3.428) \\ 
F06 & -35.1 (1.154) & -116.1 (0.810) & 120.9 (0.651) &  -8.8 (4.862) &  -3.5 (0.013) & -42.6 (4.132) \\ 
\end{tabular}
\caption{Estimates of the mean potential energy changes (kcal/mol) upon the binding of various ligands to CB[7].
The columns refer to van der Waals (VDW), coulomb (Coul), electrostatic solvation (PB), valence (Val, bond + angle + dihedral), nonpolar solvation (NP), and total energies.
The value in the parentheses is the standard deviation from bootstrapping: the observable is estimated based on 1000 random selections of 100 values of $\hat{\Theta}$.  
In Table \ref{tab:S2} of the Supplemental Material, mean potential energies for the ligand, receptor, and complex are also shown.
\label{tab:PE}}
\end{table*}

Compared to the full procedure for estimating the binding PMF, applying the dominant state configuration leads to a reduction in the correlation with M2 results and an increase in the RMSE (Table \ref{tab:FE_approx}).  In contrast, applying the dominant state approximation to calculate $\Delta \hat{G}^\circ$ from $\hat{B}(r_R)$ leads to a near-constant reduction of about 3 kcal/mol in the estimated binding free energy.  While the RMSE increases, the correlation with M2 results remains nearly identical.  Given the same $\hat{B}(r_R)$ results, however, there is essentially no reason to apply the dominant state approximation rather than Eq. (\ref{eq:imLigFE_Isample}).

\begin{table*}
\center
\begin{tabular}{l | c c c c }
Ligand & & & & \\ 
\hline
$\hat{B}(r_R)$ & min$\left\{\Psi(r_R)\right\}$ & min$\left\{\Psi(r_R)\right\}$ & HREX & HREX \\ 
$\Delta \hat{G}^\circ$ & min$\left\{\hat{B}(r_R)\right\}$ & EXP & min$\left\{\hat{B}(r_R)\right\}$ & EXP \\ 
\hline
AD1 & -28.6 & -27.2 & -22.0 & -20.1 \\ 
AD2 & -36.4 & -34.6 & -27.6 & -25.4 \\ 
AD3 & -38.1 & -36.8 & -27.6 & -26.2 \\ 
AD4 & -43.1 & -40.4 & -29.8 & -27.1 \\ 
AD5 & -35.8 & -33.6 & -26.8 & -24.4 \\ 
B02 & -29.8 & -27.9 & -21.0 & -18.1 \\ 
B05 & -37.9 & -35.6 & -23.7 & -21.4 \\ 
B11 & -48.5 & -45.7 & -23.1 & -20.5 \\ 
F01 & -22.7 & -21.3 & -10.2 &  -7.6 \\ 
F02 & -30.9 & -28.8 & -17.0 & -14.6 \\ 
F03 & -28.7 & -27.0 & -14.5 & -13.2 \\ 
F06 & -35.6 & -33.8 & -21.3 & -19.7 \\ 
\hline
R$^2_{ITC}$ & 0.849 & 0.855 & 0.684 & 0.704 \\ 
RMSE$_{ITC}$ &  17.3 &  15.3 &   5.8 &   4.5 \\ 
\hline
R$^2_{Gilson}$ & 0.787 & 0.795 & 0.926 & 0.925 \\ 
RMSE$_{Gilson}$ &  15.8 &  13.9 &   3.5 &   2.4 \\ 
\hline
R$^2_{Exp}$ & 0.723 & 0.736 & 0.996 & \\ 
RMSE$_{Exp}$ &  15.5 &  13.6 &   2.3 & \\ 
\end{tabular}
\caption{Estimates of the binding free energy $\Delta G^\circ$ (kcal/mol) using the PBSA model.  
First, the binding PMF $B(r_R)$ is estimated with the dominant state approximation ($\min\left\{\Psi(r_R)\right\}$) or based on Eq. (\ref{eq:bindingPMF_decomposition}) (HREX).  
Then, $\Delta \hat{G}^\circ$ is from the dominant state approximation ($\min\left\{\hat{B}(r_R)\right\}$) or based on 
Eq. (\ref{eq:imLigFE_Isample}) (EXP).  
The bottom rows show the correlation coefficient (R$^2$) and root mean square error (RMSE, Eq. (\ref{eq:RMSE})) with respect to isothermal titration calorimetry experiments (ITC) and mining minima calculations (Gilson) from \citet{Moghaddam2011}, and the fourth column.\label{tab:FE_approx}}
\end{table*}

There is considerable variation in the binding PMFs for the 100 receptor structures (Fig. \ref{fig:B02} and Fig. \ref{fig:S2} in the Supplemental Material.  For most of the ligands, the range of binding PMFs span 10 to 20 kcal/mol.  
While the binding PMF of the minimized structure is often near the lower end of the binding PMF distribution, this is not always the case.
In larger ligands, the binding PMF appears to be lower for other receptor structures.
The fact that a single structure does not always lead to the lowest binding PMF shows a major limitation of using a single receptor structure to estimate binding free energies.

\begin{figure}
\includegraphics{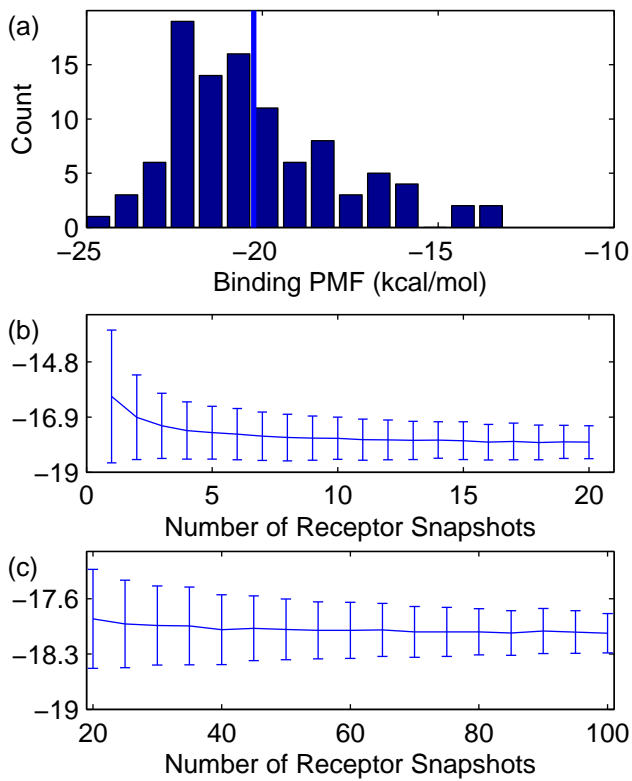}
\caption{
(a) Histogram of binding PMF estimates $\hat{B}(r_R)$ (kcal/mol) of B02 to 100 snapshots of CB[7], using PBSA energies.  The vertical line shows the mean binding PMF for the minimized receptor structure.
(b) and (c) Estimates of the binding free energy $\Delta G^\circ$ of B02 to CB[7] (kcal/mol), using PBSA energies, as a function of the number of receptor snapshots.
The line and error bars denote the mean and standard deviation from bootstrapping: the binding free energy is estimated 100 times using random selections of $N$ out of 100 binding PMFs.
Analogous plots for the other ligands in this study are available as Figs. \ref{fig:S2} and \ref{fig:S3} in the Supplemental Material.
\label{fig:B02}}
\end{figure}

In spite of the variability of binding PMFs, for the ligands in the test set, the average value of $\Delta G^\circ$ appears to stabilize after a relatively small number (about 15) of receptor snapshots (Fig. \ref{fig:B02} and Fig. S3 in the Supplemental Material.  Using a greater number of snapshots slightly reduces the variance of binding free energy estimates.  After the certain point, however, further reduction in the variance of $\Delta \hat{G}^\circ$ is limited by the variance in binding PMF estimates.

\section{Discussion}

While the good agreement between implicit ligand and M2 calculations provides a proof of principle, the convergence and accuracy of implicit ligand calculations will differ with other classes of receptor-ligand pairs.
With protein-ligand pairs, for example, representative sampling of receptors and finding low-energy poses of the ligand will likely require much more MD simulation time.
On the other hand, many protein-ligand systems are not as strongly charged and may be less sensitive to the electrostatic solvation free energy.
Due to these variabilities, assessments for the feasibility of implicit ligand calculations in different classes of systems will prove valuable.
Tests for convergence and accuracy may be similar to those performed for the CB[7] system.

Numerous opportunities remain for further methodological improvement and optimization of implicit ligand free energy calculations.
The accuracy of implicit ligand calculations (and M2 calculations) may be limited by the quality of the force field.
The decomposition of the binding PMF in Eq. (\ref{eq:bindingPMF_decomposition}) provides a facile means to integrate alternate and potentially more expensive potential energies, e.g. quantum mechanical calculations or more sophisticated nonpolar solvation free energies.
Modeling may also be improved by the inclusion of a few explicit water molecules (see Supplemental Material.)
Another potential avenue for improvement is the fine-tuning of the replica exchange protocol (e.g. using implicit solvent or optimizing the number of stages and values of $\lambda$ for a particular system) or implementing alternative methods to estimate the binding PMF and binding free energy.

Even without modifying the replica exchange protocol, computations may be accelerated by optimizing existing MD simulation packages for implicit ligand theory.  Few modern MD simulation programs take full advantage of rigid degrees of freedom by skipping the calculation of pairwise interactions between rigid atoms.  Even fewer implicit solvent models are designed with rigid receptors in mind \cite{Guvench2001}; implicit ligand theory may inspire the development of such models.

Implicit ligand theory also provides guidance on how to understand and improve existing molecular docking algorithms.
The definition of $\Psi(r_{RL})$ provides a straightforward functional form that can be used to account for solvation free energies and ligand internal energies (strain), which have been noted to be important factors in binding free energies \cite{Mobley2009}, but are frequently ignored in the interaction energy functions used by docking packages.
Implicit ligand theory also delineates how to improve the ranking of different ligands.
Molecular docking packages currently rank receptor-ligand binding free energies based on a single low-energy configuration.
As such, they apply the crudest form of implicit ligand theory, the dominant state approximation, to estimate both the binding PMF and the binding free energy.  
With important modifications to existing algorithms and the application of more complex estimators, the accuracy of scoring functions should be enhanced.

One important potential change to molecular docking is the inclusion of multiple receptor configurations.
While most modern docking packages account for the orientation and flexibility of the ligand, the large number of coordinates makes the treatment of receptor flexibility challenging.
A number of groups have improved docking performance by treating receptor flexibility by using multiple structures from crystallography \cite{Rao:2008p6674, Craig:2010p6676, Bottegoni:2011p6687} or MD simulations \cite{Nichols:2011p6682} (the \emph{relaxed complex method} \cite{Lin:2002p6859,Lin:2003p6839,Amaro:2008p6685}).
Molecular dynamics simulations have also revealed binding sites not discovered by crystallography \cite{Schames:2004p6724,Amaro:2007p2}.  In the case of HIV integrase, insight into a new binding site even inspired the development of a new drug \cite{Durrant:2011p6673}.

Despite of this success, it has hitherto remained unclear how to combine information from docking to different receptor snapshots.  
While averaging strategies have been empirically compared \cite{Paulsen2009}, the default strategy has been to rank the ligand using the minimal energy from docking to all the snapshots.  
With implicit ligand theory, it is clear that the binding free energy may also be estimated by using an exponential average or cumulant expansion of the binding PMF (which may still come from the dominant state approximation) for different snapshots.

The computational expense of the relaxed complex approach may be reduced by clustering snapshots and selecting a representative snapshot from each cluster \cite{Amaro:2008p6685}.  
Assuming that the binding PMF is constant within the cluster, estimated averages may be weighed by the cluster size.
Using a clustering algorithm based on QR factorization to select 33 representative structures, \citet{Amaro:2008p6685} were able to accurately reproduce a histogram of docking scores to over 400 structures.  
Further research will be necessary to develop and validate algorithms that reliably cluster receptor configurations in which the binding PMF is nearly constant.

Compared to the inclusion of multiple receptor structures, a more difficult task is the estimation of $B(r_R)$ using molecular docking, as this involves a paradigm shift from searching for a minimum to sampling from a distribution.  Docking algorithms may be broadly classified into two categories: matching and docking simulation \cite{Morris:1998p6664}.  Matching algorithms such as DOCK \cite{KUNTZ:1982p6867} attempt to match a ligand into a model of the binding site.  DOCK models both the ligand and the binding site as a set of spheres and uses algorithms from graph theory to align the ligand spheres into the binding site spheres.  In docking simulation methods such as AutoDOCK \cite{Goodsell:1990p6732, Morris:1996p6733, Morris:1998p6664, Huey:2007p6666} and MCDOCK \cite{Liu:1999p6838}, the ligand starts outside the binding site and its configuration and orientation are progressively modified to search for the lowest energy configuration of the complex.

Matching algorithms can estimate $B(r_R)$ by a postprocessing algorithm.  That is, after low-energy complexes are found, they may be used to bias receptor-independent random sampling of the ligand orientation by a confining potential $U_c(\xi_L)$, for use in Eq. (\ref{eq:confinedBindingPMF}).
With a harmonic potential for $U_c(\xi_L)$, the ligand orientation will come from a Gaussian distribution.
An alternative postprocessing algorithm is to use the lowest-energy structure from a matching algorithm as a starting point for a rigid-receptor MD simulation.
This is not prohibitively expensive; \citet{Graves:2008p6534} even used MD simulations with  flexibility near the binding site as a postprocessing step for molecular docking.  Samples from this simulation would be used to estimate $B(r_R)$ based on Eq. (\ref{eq:IbindingPMF}) or Eq. (\ref{eq:IconfinedBindingPMF}).

Docking simulation methods, on the other hand, will need to be modified to sample from a known distribution rather than to search for the minimum energy.  This change may not require a complete revamp.  Docking simulation algorithms are often based on Monte Carlo approaches, which preserve a desired distribution or may be readily modified to do so.  For example, MCDOCK \cite{Liu:1999p6838} and early generations of AutoDOCK \cite{Goodsell:1990p6732, Morris:1996p6733} use simulated annealing, a procedure for which it is possible to calculate the importance sampling weight \cite{Neal:2001p3154}.

In addition to providing a path to rigorous binding free energies from molecular docking, implicit ligand theory also quantifies existing notions \cite{Carlson:2002p6715} about whether molecular recognition proceeds by induced fit or conformational selection \cite{Xu:2008p6806, Bucher:2011p6810}.  As all receptor configurations have finite Boltzmann probability, the issue is a matter of degree.  Suppose that a receptor binds to two different ligands with the same binding free energy, one by conformational selection and the other by induced fit.  If, to a good approximation, the complex is dominated by a single structure with receptor configuration $r_R^*$ such that $B(r_R)=\infty$ for all other receptor configurations, then Eq. (\ref{eq:FE_implicitL}) simplifies to $\Delta G^\circ = \U(r_R^*) + B(r_R^*) + \beta^{-1} \ln Z_R + \Delta  G_\xi$.  For the ligand that binds by conformational selection, $p(r_R^*) = e^{-\beta \U(r_R^*)}/Z_R$ has a reasonably high probability.  In the induced fit complex, $\U(r_R^*)$ is much less favorable and $B(r_R^*)$ must compensate accordingly to achieve the same $\Delta G^\circ$.

Source code and data used in this paper are available at \url{https://simtk.org/home/implicit_ligand}.

\section{Acknowledgement}

The author thanks David Beratan for being a supportive postdoctoral advisor,
Aaron Virshup and Shahar Keinan for helpful discussions,
John Chodera and David Mobley for comments on the manuscript, 
Yi Wang for suggesting CB[7] as a test case,
Michael Gilson for providing parameters for CB[7] and its ligands, 
Clayton Jarratt for pinpointing the cause of negative surface areas in NAMD,
and Emilio Gallichio for sharing source code for BEDAM.
Calculations were performed using Duke Shared Computing Resources (DSCR).
This research was funded by NSF CHE10-57953, NIH 2P50 GM-067082-06-10, and N00014-11-1-0729.


\begin{thebibliography}{61}%
\makeatletter
\providecommand \@ifxundefined [1]{%
 \@ifx{#1\undefined}
}%
\providecommand \@ifnum [1]{%
 \ifnum #1\expandafter \@firstoftwo
 \else \expandafter \@secondoftwo
 \fi
}%
\providecommand \@ifx [1]{%
 \ifx #1\expandafter \@firstoftwo
 \else \expandafter \@secondoftwo
 \fi
}%
\providecommand \natexlab [1]{#1}%
\providecommand \enquote  [1]{``#1''}%
\providecommand \bibnamefont  [1]{#1}%
\providecommand \bibfnamefont [1]{#1}%
\providecommand \citenamefont [1]{#1}%
\providecommand \href@noop [0]{\@secondoftwo}%
\providecommand \href [0]{\begingroup \@sanitize@url \@href}%
\providecommand \@href[1]{\@@startlink{#1}\@@href}%
\providecommand \@@href[1]{\endgroup#1\@@endlink}%
\providecommand \@sanitize@url [0]{\catcode `\\12\catcode `\$12\catcode
  `\&12\catcode `\#12\catcode `\^12\catcode `\_12\catcode `\%12\relax}%
\providecommand \@@startlink[1]{}%
\providecommand \@@endlink[0]{}%
\providecommand \url  [0]{\begingroup\@sanitize@url \@url }%
\providecommand \@url [1]{\endgroup\@href {#1}{\urlprefix }}%
\providecommand \urlprefix  [0]{URL }%
\providecommand \Eprint [0]{\href }%
\providecommand \doibase [0]{http://dx.doi.org/}%
\providecommand \selectlanguage [0]{\@gobble}%
\providecommand \bibinfo  [0]{\@secondoftwo}%
\providecommand \bibfield  [0]{\@secondoftwo}%
\providecommand \translation [1]{[#1]}%
\providecommand \BibitemOpen [0]{}%
\providecommand \bibitemStop [0]{}%
\providecommand \bibitemNoStop [0]{.\EOS\space}%
\providecommand \EOS [0]{\spacefactor3000\relax}%
\providecommand \BibitemShut  [1]{\csname bibitem#1\endcsname}%
\let\auto@bib@innerbib\@empty
\bibitem [{\citenamefont {Shoichet}(2004)}]{Shoichet:2004p6935}%
  \BibitemOpen
  \bibfield  {author} {\bibinfo {author} {\bibfnamefont {B.~K.}\ \bibnamefont
  {Shoichet}},\ }\href {\doibase 10.1038/nature03197} {\bibfield  {journal}
  {\bibinfo  {journal} {Nature}\ }\textbf {\bibinfo {volume} {432}},\ \bibinfo
  {pages} {862} (\bibinfo {year} {2004})}\BibitemShut {NoStop}%
\bibitem [{\citenamefont {Klebe}(2006)}]{Klebe:2006p6699}%
  \BibitemOpen
  \bibfield  {author} {\bibinfo {author} {\bibfnamefont {G.}~\bibnamefont
  {Klebe}},\ }\href {\doibase 10.1016/j.drudis.2006.05.012} {\bibfield
  {journal} {\bibinfo  {journal} {Drug Discov Today}\ }\textbf {\bibinfo
  {volume} {11}},\ \bibinfo {pages} {580} (\bibinfo {year} {2006})}\BibitemShut
  {NoStop}%
\bibitem [{\citenamefont {Kim}\ and\ \citenamefont
  {Skolnick}(2008)}]{Kim:2008p6686}%
  \BibitemOpen
  \bibfield  {author} {\bibinfo {author} {\bibfnamefont {R.}~\bibnamefont
  {Kim}}\ and\ \bibinfo {author} {\bibfnamefont {J.}~\bibnamefont {Skolnick}},\
  }\href {\doibase 10.1002/jcc.20893} {\bibfield  {journal} {\bibinfo
  {journal} {J Comput Chem}\ }\textbf {\bibinfo {volume} {29}},\ \bibinfo
  {pages} {1316} (\bibinfo {year} {2008})}\BibitemShut {NoStop}%
\bibitem [{\citenamefont {Moitessier}\ \emph {et~al.}(2009)\citenamefont
  {Moitessier}, \citenamefont {Englebienne}, \citenamefont {Lee}, \citenamefont
  {Lawandi},\ and\ \citenamefont {Corbeil}}]{Moitessier:2009p6716}%
  \BibitemOpen
  \bibfield  {author} {\bibinfo {author} {\bibfnamefont {N.}~\bibnamefont
  {Moitessier}}, \bibinfo {author} {\bibfnamefont {P.}~\bibnamefont
  {Englebienne}}, \bibinfo {author} {\bibfnamefont {D.}~\bibnamefont {Lee}},
  \bibinfo {author} {\bibfnamefont {J.}~\bibnamefont {Lawandi}}, \ and\
  \bibinfo {author} {\bibfnamefont {C.~R.}\ \bibnamefont {Corbeil}},\ }\href
  {\doibase 10.1038/sj.bjp.0707515} {\bibfield  {journal} {\bibinfo  {journal}
  {British Journal of Pharmacology}\ }\textbf {\bibinfo {volume} {153}},\
  \bibinfo {pages} {S7} (\bibinfo {year} {2009})}\BibitemShut {NoStop}%
\bibitem [{\citenamefont {Plewczynski}\ \emph {et~al.}(2010)\citenamefont
  {Plewczynski}, \citenamefont {La{\'z}niewski}, \citenamefont {Augustyniak},\
  and\ \citenamefont {Ginalski}}]{Plewczynski:2010p6700}%
  \BibitemOpen
  \bibfield  {author} {\bibinfo {author} {\bibfnamefont {D.}~\bibnamefont
  {Plewczynski}}, \bibinfo {author} {\bibfnamefont {M.}~\bibnamefont
  {La{\'z}niewski}}, \bibinfo {author} {\bibfnamefont {R.}~\bibnamefont
  {Augustyniak}}, \ and\ \bibinfo {author} {\bibfnamefont {K.}~\bibnamefont
  {Ginalski}},\ }\href {\doibase 10.1002/jcc.21643} {\bibfield  {journal}
  {\bibinfo  {journal} {J Comput Chem}\ }\textbf {\bibinfo {volume} {32}},\
  \bibinfo {pages} {742} (\bibinfo {year} {2010})}\BibitemShut {NoStop}%
\bibitem [{\citenamefont {Wang}\ \emph {et~al.}(2001)\citenamefont {Wang},
  \citenamefont {Morin}, \citenamefont {Wang},\ and\ \citenamefont
  {Kollman}}]{Wang:2001p7494}%
  \BibitemOpen
  \bibfield  {author} {\bibinfo {author} {\bibfnamefont {J.}~\bibnamefont
  {Wang}}, \bibinfo {author} {\bibfnamefont {P.}~\bibnamefont {Morin}},
  \bibinfo {author} {\bibfnamefont {W.}~\bibnamefont {Wang}}, \ and\ \bibinfo
  {author} {\bibfnamefont {P.}~\bibnamefont {Kollman}},\ }\href {\doibase
  10.1021/ja003834q} {\bibfield  {journal} {\bibinfo  {journal} {J Am Chem
  Soc}\ }\textbf {\bibinfo {volume} {123}},\ \bibinfo {pages} {5221} (\bibinfo
  {year} {2001})}\BibitemShut {NoStop}%
\bibitem [{\citenamefont {Lin}\ \emph {et~al.}(2003)\citenamefont {Lin},
  \citenamefont {Perryman}, \citenamefont {Schames},\ and\ \citenamefont
  {McCammon}}]{Lin:2003p6839}%
  \BibitemOpen
  \bibfield  {author} {\bibinfo {author} {\bibfnamefont {J.}~\bibnamefont
  {Lin}}, \bibinfo {author} {\bibfnamefont {A.}~\bibnamefont {Perryman}},
  \bibinfo {author} {\bibfnamefont {J.}~\bibnamefont {Schames}}, \ and\
  \bibinfo {author} {\bibfnamefont {J.}~\bibnamefont {McCammon}},\ }\href
  {\doibase 10.1002/bip.10218} {\bibfield  {journal} {\bibinfo  {journal}
  {Biopolymers}\ }\textbf {\bibinfo {volume} {68}},\ \bibinfo {pages} {47}
  (\bibinfo {year} {2003})}\BibitemShut {NoStop}%
\bibitem [{\citenamefont {Graves}\ \emph {et~al.}(2008)\citenamefont {Graves},
  \citenamefont {Shivakumar}, \citenamefont {Boyce}, \citenamefont {Jacobson},
  \citenamefont {Case},\ and\ \citenamefont {Shoichet}}]{Graves:2008p6534}%
  \BibitemOpen
  \bibfield  {author} {\bibinfo {author} {\bibfnamefont {A.~P.}\ \bibnamefont
  {Graves}}, \bibinfo {author} {\bibfnamefont {D.~M.}\ \bibnamefont
  {Shivakumar}}, \bibinfo {author} {\bibfnamefont {S.~E.}\ \bibnamefont
  {Boyce}}, \bibinfo {author} {\bibfnamefont {M.~P.}\ \bibnamefont {Jacobson}},
  \bibinfo {author} {\bibfnamefont {D.~A.}\ \bibnamefont {Case}}, \ and\
  \bibinfo {author} {\bibfnamefont {B.~K.}\ \bibnamefont {Shoichet}},\ }\href
  {\doibase 10.1016/j.jmb.2008.01.049} {\bibfield  {journal} {\bibinfo
  {journal} {J Mol Biol}\ }\textbf {\bibinfo {volume} {377}},\ \bibinfo {pages}
  {914} (\bibinfo {year} {2008})}\BibitemShut {NoStop}%
\bibitem [{\citenamefont {Thompson}\ \emph {et~al.}(2008)\citenamefont
  {Thompson}, \citenamefont {Humblet},\ and\ \citenamefont
  {Joseph-McCarthy}}]{Thompson:2008p7384}%
  \BibitemOpen
  \bibfield  {author} {\bibinfo {author} {\bibfnamefont {D.~C.}\ \bibnamefont
  {Thompson}}, \bibinfo {author} {\bibfnamefont {C.}~\bibnamefont {Humblet}}, \
  and\ \bibinfo {author} {\bibfnamefont {D.}~\bibnamefont {Joseph-McCarthy}},\
  }\href {\doibase 10.1021/ci700470c} {\bibfield  {journal} {\bibinfo
  {journal} {J Chem Inf Model}\ }\textbf {\bibinfo {volume} {48}},\ \bibinfo
  {pages} {1081} (\bibinfo {year} {2008})}\BibitemShut {NoStop}%
\bibitem [{\citenamefont {Hou}\ \emph {et~al.}(2010)\citenamefont {Hou},
  \citenamefont {Wang}, \citenamefont {Li},\ and\ \citenamefont
  {Wang}}]{Hou:2010p6692}%
  \BibitemOpen
  \bibfield  {author} {\bibinfo {author} {\bibfnamefont {T.}~\bibnamefont
  {Hou}}, \bibinfo {author} {\bibfnamefont {J.}~\bibnamefont {Wang}}, \bibinfo
  {author} {\bibfnamefont {Y.}~\bibnamefont {Li}}, \ and\ \bibinfo {author}
  {\bibfnamefont {W.}~\bibnamefont {Wang}},\ }\href {\doibase
  10.1002/jcc.21666} {\bibfield  {journal} {\bibinfo  {journal} {J Comput
  Chem}\ }\textbf {\bibinfo {volume} {32}},\ \bibinfo {pages} {866} (\bibinfo
  {year} {2010})}\BibitemShut {NoStop}%
\bibitem [{\citenamefont {Chang}\ \emph {et~al.}(2010)\citenamefont {Chang},
  \citenamefont {Ayeni}, \citenamefont {Breuer},\ and\ \citenamefont
  {Torbett}}]{Chang:2010p7063}%
  \BibitemOpen
  \bibfield  {author} {\bibinfo {author} {\bibfnamefont {M.~W.}\ \bibnamefont
  {Chang}}, \bibinfo {author} {\bibfnamefont {C.}~\bibnamefont {Ayeni}},
  \bibinfo {author} {\bibfnamefont {S.}~\bibnamefont {Breuer}}, \ and\ \bibinfo
  {author} {\bibfnamefont {B.~E.}\ \bibnamefont {Torbett}},\ }\href {\doibase
  10.1371/journal.pone.0011955.t002} {\bibfield  {journal} {\bibinfo  {journal}
  {PLoS ONE}\ }\textbf {\bibinfo {volume} {5}},\ \bibinfo {pages} {e11955}
  (\bibinfo {year} {2010})}\BibitemShut {NoStop}%
\bibitem [{\citenamefont {Huang}\ \emph {et~al.}(2006)\citenamefont {Huang},
  \citenamefont {Kalyanaraman}, \citenamefont {Bernacki},\ and\ \citenamefont
  {Jacobson}}]{Huang:2006p6690}%
  \BibitemOpen
  \bibfield  {author} {\bibinfo {author} {\bibfnamefont {N.}~\bibnamefont
  {Huang}}, \bibinfo {author} {\bibfnamefont {C.}~\bibnamefont {Kalyanaraman}},
  \bibinfo {author} {\bibfnamefont {K.}~\bibnamefont {Bernacki}}, \ and\
  \bibinfo {author} {\bibfnamefont {M.~P.}\ \bibnamefont {Jacobson}},\ }\href
  {\doibase 10.1039/b608269f} {\bibfield  {journal} {\bibinfo  {journal} {Phys
  Chem Chem Phys}\ }\textbf {\bibinfo {volume} {8}},\ \bibinfo {pages}
  {5166} (\bibinfo {year} {2006})}\BibitemShut {NoStop}%
\bibitem [{Note1()}]{Note1}%
  \BibitemOpen
  \bibinfo {note} {Activities have been assumed to be unity, a reasonable
  approximation in the limit of low concentrations.}\BibitemShut {Stop}%
\bibitem [{\citenamefont {Gilson}\ \emph {et~al.}(1997)\citenamefont {Gilson},
  \citenamefont {Given}, \citenamefont {Bush},\ and\ \citenamefont
  {McCammon}}]{Gilson:1997p6409}%
  \BibitemOpen
  \bibfield  {author} {\bibinfo {author} {\bibfnamefont {M.~K.}\ \bibnamefont
  {Gilson}}, \bibinfo {author} {\bibfnamefont {J.}~\bibnamefont {Given}},
  \bibinfo {author} {\bibfnamefont {B.}~\bibnamefont {Bush}}, \ and\ \bibinfo
  {author} {\bibfnamefont {J.}~\bibnamefont {McCammon}},\ }\href@noop {}
  {\bibfield  {journal} {\bibinfo  {journal} {Biophys J}\ }\textbf {\bibinfo
  {volume} {72}},\ \bibinfo {pages} {1047} (\bibinfo {year}
  {1997})}\BibitemShut {NoStop}%
\bibitem [{\citenamefont {Wang}\ \emph {et~al.}(2008)\citenamefont {Wang},
  \citenamefont {Tan}, \citenamefont {Tan}, \citenamefont {Lu},\ and\
  \citenamefont {Luo}}]{Wang:2008p6628}%
  \BibitemOpen
  \bibfield  {author} {\bibinfo {author} {\bibfnamefont {J.}~\bibnamefont
  {Wang}}, \bibinfo {author} {\bibfnamefont {C.}~\bibnamefont {Tan}}, \bibinfo
  {author} {\bibfnamefont {Y.-H.}\ \bibnamefont {Tan}}, \bibinfo {author}
  {\bibfnamefont {Q.}~\bibnamefont {Lu}}, \ and\ \bibinfo {author}
  {\bibfnamefont {R.}~\bibnamefont {Luo}},\ }\href@noop {} {\bibfield
  {journal} {\bibinfo  {journal} {Commun Comput Phys}\ }\textbf {\bibinfo
  {volume} {3}},\ \bibinfo {pages} {1010} (\bibinfo {year} {2008})}\BibitemShut
  {NoStop}%
\bibitem [{\citenamefont {Feig}\ and\ \citenamefont
  {Brooks}(2004)}]{Feig:2004p6763}%
  \BibitemOpen
  \bibfield  {author} {\bibinfo {author} {\bibfnamefont {M.}~\bibnamefont
  {Feig}}\ and\ \bibinfo {author} {\bibfnamefont {C.}~\bibnamefont {Brooks}},\
  }\href {\doibase 10.1016/j.sbi.2004.03.009} {\bibfield  {journal} {\bibinfo
  {journal} {Curr Opin Struc Biol}\ }\textbf {\bibinfo {volume} {14}},\
  \bibinfo {pages} {217} (\bibinfo {year} {2004})}\BibitemShut {NoStop}%
\bibitem [{\citenamefont {Michel}\ and\ \citenamefont
  {Essex}(2008)}]{Michel2008}%
  \BibitemOpen
  \bibfield  {author} {\bibinfo {author} {\bibfnamefont {J.}~\bibnamefont
  {Michel}}\ and\ \bibinfo {author} {\bibfnamefont {J.~W.}\ \bibnamefont
  {Essex}},\ }\href {\doibase 10.1021/jm800524s} {\bibfield  {journal}
  {\bibinfo  {journal} {J Med Chem}\ }\textbf {\bibinfo
  {volume} {51}},\ \bibinfo {pages} {6654} (\bibinfo {year}
  {2008})}\BibitemShut {NoStop}%
\bibitem [{\citenamefont {Chang}\ and\ \citenamefont
  {Gilson}(2003)}]{Chang2003}%
  \BibitemOpen
  \bibfield  {author} {\bibinfo {author} {\bibfnamefont {C.-E.}\ \bibnamefont
  {Chang}}\ and\ \bibinfo {author} {\bibfnamefont {M.~K.}\ \bibnamefont
  {Gilson}},\ }\href {\doibase 10.1002/jcc.10325} {\bibfield  {journal}
  {\bibinfo  {journal} {J Comput Chem}\ }\textbf {\bibinfo
  {volume} {24}},\ \bibinfo {pages} {1987} (\bibinfo {year}
  {2003})}\BibitemShut {NoStop}%
\bibitem [{\citenamefont {Chang}\ \emph {et~al.}(2003)\citenamefont {Chang},
  \citenamefont {Potter},\ and\ \citenamefont {Gilson}}]{Chang2003a}%
  \BibitemOpen
  \bibfield  {author} {\bibinfo {author} {\bibfnamefont {C.-E.}\ \bibnamefont
  {Chang}}, \bibinfo {author} {\bibfnamefont {M.~J.}\ \bibnamefont {Potter}}, \
  and\ \bibinfo {author} {\bibfnamefont {M.~K.}\ \bibnamefont {Gilson}},\
  }\href {\doibase 10.1021/jp027149c} {\bibfield  {journal} {\bibinfo
  {journal} {The Journal of Physical Chemistry B}\ }\textbf {\bibinfo {volume}
  {107}},\ \bibinfo {pages} {1048} (\bibinfo {year} {2003})}\BibitemShut
  {NoStop}%
\bibitem [{\citenamefont {Lee}(2005)}]{Lee2005}%
  \BibitemOpen
  \bibfield  {author} {\bibinfo {author} {\bibfnamefont {M.}~\bibnamefont
  {Lee}},\ }\href {papers://f7841d7d-9771-4011-86e7-7e86ad060f2b/Paper/p3130}
  {\bibfield  {journal} {\bibinfo  {journal} {Biophys J}\ }\textbf
  {\bibinfo {volume} {90}},\ \bibinfo {pages} {864} (\bibinfo {year}
  {2005})}\BibitemShut {NoStop}%
\bibitem [{\citenamefont {Gallicchio}\ \emph {et~al.}(2010)\citenamefont
  {Gallicchio}, \citenamefont {Lapelosa},\ and\ \citenamefont
  {Levy}}]{Gallicchio2010}%
  \BibitemOpen
  \bibfield  {author} {\bibinfo {author} {\bibfnamefont {E.}~\bibnamefont
  {Gallicchio}}, \bibinfo {author} {\bibfnamefont {M.}~\bibnamefont
  {Lapelosa}}, \ and\ \bibinfo {author} {\bibfnamefont {R.~M.}\ \bibnamefont
  {Levy}},\ }\href {http://pubs.acs.org/doi/abs/10.1021/ct1002913
  papers://f7841d7d-9771-4011-86e7-7e86ad060f2b/Paper/p7963} {\bibfield
  {journal} {\bibinfo  {journal} {J Chem Theory Comput}\
  }\textbf {\bibinfo {volume} {6}},\ \bibinfo {pages} {2961} (\bibinfo {year}
  {2010})}\BibitemShut {NoStop}%
\bibitem [{\citenamefont {Cohen}\ \emph {et~al.}(2006)\citenamefont {Cohen},
  \citenamefont {Arkhipov}, \citenamefont {Braun},\ and\ \citenamefont
  {Schulten}}]{Cohen2006}%
  \BibitemOpen
  \bibfield  {author} {\bibinfo {author} {\bibfnamefont {J.}~\bibnamefont
  {Cohen}}, \bibinfo {author} {\bibfnamefont {A.}~\bibnamefont {Arkhipov}},
  \bibinfo {author} {\bibfnamefont {R.}~\bibnamefont {Braun}}, \ and\ \bibinfo
  {author} {\bibfnamefont {K.}~\bibnamefont {Schulten}},\ }\href
  {http://dx.doi.org/10.1529/biophysj.106.085746
  papers://f7841d7d-9771-4011-86e7-7e86ad060f2b/Paper/p8331} {\bibfield
  {journal} {\bibinfo  {journal} {Biophys J}\ }\textbf {\bibinfo
  {volume} {91}},\ \bibinfo {pages} {1844} (\bibinfo {year}
  {2006})}\BibitemShut {NoStop}%
\bibitem [{\citenamefont {Jiang}\ \emph {et~al.}(2009)\citenamefont {Jiang},
  \citenamefont {Hodoscek},\ and\ \citenamefont {Roux}}]{Jiang2009}%
  \BibitemOpen
  \bibfield  {author} {\bibinfo {author} {\bibfnamefont {W.}~\bibnamefont
  {Jiang}}, \bibinfo {author} {\bibfnamefont {M.}~\bibnamefont {Hodoscek}}, \
  and\ \bibinfo {author} {\bibfnamefont {B.}~\bibnamefont {Roux}},\ }\href
  {http://pubs.acs.org/doi/abs/10.1021/ct900223z
  papers://f7841d7d-9771-4011-86e7-7e86ad060f2b/Paper/p6598} {\bibfield
  {journal} {\bibinfo  {journal} {J Chem Theory Comput}\
  }\textbf {\bibinfo {volume} {5}},\ \bibinfo {pages} {2583} (\bibinfo {year}
  {2009})}\BibitemShut {NoStop}%
\bibitem [{\citenamefont {Gallicchio}\ and\ \citenamefont
  {Levy}(2012)}]{Gallicchio2012}%
  \BibitemOpen
  \bibfield  {author} {\bibinfo {author} {\bibfnamefont {E.}~\bibnamefont
  {Gallicchio}}\ and\ \bibinfo {author} {\bibfnamefont {R.~M.}\ \bibnamefont
  {Levy}},\ }\href {papers://f7841d7d-9771-4011-86e7-7e86ad060f2b/Paper/p8054}
  {\bibfield  {journal} {\bibinfo  {journal} {J Comput-Aided Mol Des}\ ,\ \bibinfo {pages} {1}} (\bibinfo {year}
  {2012})}\BibitemShut {NoStop}%
\bibitem [{\citenamefont {Rao}\ \emph {et~al.}(2008)\citenamefont {Rao},
  \citenamefont {Sanschagrin}, \citenamefont {Greenwood}, \citenamefont
  {Repasky}, \citenamefont {Sherman},\ and\ \citenamefont
  {Farid}}]{Rao:2008p6674}%
  \BibitemOpen
  \bibfield  {author} {\bibinfo {author} {\bibfnamefont {S.}~\bibnamefont
  {Rao}}, \bibinfo {author} {\bibfnamefont {P.~C.}\ \bibnamefont
  {Sanschagrin}}, \bibinfo {author} {\bibfnamefont {J.~R.}\ \bibnamefont
  {Greenwood}}, \bibinfo {author} {\bibfnamefont {M.~P.}\ \bibnamefont
  {Repasky}}, \bibinfo {author} {\bibfnamefont {W.}~\bibnamefont {Sherman}}, \
  and\ \bibinfo {author} {\bibfnamefont {R.}~\bibnamefont {Farid}},\ }\href
  {\doibase 10.1007/s10822-008-9182-y} {\bibfield  {journal} {\bibinfo
  {journal} {J Comput Aided Mol Des}\ }\textbf {\bibinfo {volume} {22}},\
  \bibinfo {pages} {621} (\bibinfo {year} {2008})}\BibitemShut {NoStop}%
\bibitem [{\citenamefont {Chipot}\ and\ \citenamefont
  {Pohorille}(2007)}]{Chipot2007}%
  \BibitemOpen
  \bibinfo {editor} {\bibfnamefont {C.}~\bibnamefont {Chipot}}\ and\ \bibinfo
  {editor} {\bibfnamefont {A.}~\bibnamefont {Pohorille}},\ eds.,\ \href@noop {}
  {\emph {\bibinfo {title} {Free Energy Calculations}}},\ Vol.~\bibinfo
  {volume} {86}\ (\bibinfo  {publisher} {Springer, Berlin},\ \bibinfo {year}
  {2007})\BibitemShut {NoStop}%
\bibitem [{\citenamefont {Zwanzig}(1954{\natexlab{a}})}]{Zwanzig1954}%
  \BibitemOpen
  \bibfield  {author} {\bibinfo {author} {\bibfnamefont {R.}~\bibnamefont
  {Zwanzig}},\ }\href
  {papers://f7841d7d-9771-4011-86e7-7e86ad060f2b/Paper/p6113} {\bibfield
  {journal} {\bibinfo  {journal} {J Chem Phys}\ }\textbf
  {\bibinfo {volume} {22}},\ \bibinfo {pages} {1420} (\bibinfo {year}
  {1954}{\natexlab{a}})}\BibitemShut {NoStop}%
\bibitem [{\citenamefont {Kirkwood}(1935)}]{Kirkwood1935}%
  \BibitemOpen
  \bibfield  {author} {\bibinfo {author} {\bibfnamefont {J.~G.}\ \bibnamefont
  {Kirkwood}},\ }\href
  {http://apps.webofknowledge.com/InboundService.do?SID=2FoOLNp6PbDoOGkko7M\&product=WOS\&UT=000201218900011\&SrcApp=CR\&DestFail=http\%253A\%252F\%252Fwww.webofknowledge.com\&Init=Yes\&action=retrieve\&Func=Frame\&customersID=mekentosj\&SrcAuth=mekentosj\&IsProductCode=Yes\&mode=FullRecord
  papers://f7841d7d-9771-4011-86e7-7e86ad060f2b/Paper/p7507} {\bibfield
  {journal} {\bibinfo  {journal} {J Chem Phys}\ }\textbf
  {\bibinfo {volume} {3}},\ \bibinfo {pages} {300} (\bibinfo {year}
  {1935})}\BibitemShut {NoStop}%
\bibitem [{\citenamefont {Bennett}(1976)}]{BENNETT1976}%
  \BibitemOpen
  \bibfield  {author} {\bibinfo {author} {\bibfnamefont {C.~H.}\ \bibnamefont
  {Bennett}},\ }\href
  {papers://f7841d7d-9771-4011-86e7-7e86ad060f2b/Paper/p5518} {\bibfield
  {journal} {\bibinfo  {journal} {J Comput Phys}\ }\textbf
  {\bibinfo {volume} {22}},\ \bibinfo {pages} {245} (\bibinfo {year}
  {1976})}\BibitemShut {NoStop}%
\bibitem [{\citenamefont {Zwanzig}(1954{\natexlab{b}})}]{Zwanzig:1954p6113}%
  \BibitemOpen
  \bibfield  {author} {\bibinfo {author} {\bibfnamefont {R.}~\bibnamefont
  {Zwanzig}},\ }\href@noop {} {\bibfield  {journal} {\bibinfo  {journal} {J
  Chem Phys}\ }\textbf {\bibinfo {volume} {22}},\ \bibinfo {pages} {1420}
  (\bibinfo {year} {1954}{\natexlab{b}})}\BibitemShut {NoStop}%
\bibitem [{\citenamefont {Shirts}\ and\ \citenamefont
  {Chodera}(2008)}]{Shirts2008}%
  \BibitemOpen
  \bibfield  {author} {\bibinfo {author} {\bibfnamefont {M.~R.}\ \bibnamefont
  {Shirts}}\ and\ \bibinfo {author} {\bibfnamefont {J.~D.}\ \bibnamefont
  {Chodera}},\ }\href
  {papers://f7841d7d-9771-4011-86e7-7e86ad060f2b/Paper/p1443} {\bibfield
  {journal} {\bibinfo  {journal} {J Chem Phys}\ }\textbf
  {\bibinfo {volume} {129}},\ \bibinfo {pages} {124105} (\bibinfo {year}
  {2008})}\BibitemShut {NoStop}%
\bibitem [{\citenamefont {Moghaddam}\ \emph {et~al.}(2009)\citenamefont
  {Moghaddam}, \citenamefont {Inoue},\ and\ \citenamefont
  {Gilson}}]{Moghaddam2009}%
  \BibitemOpen
  \bibfield  {author} {\bibinfo {author} {\bibfnamefont {S.}~\bibnamefont
  {Moghaddam}}, \bibinfo {author} {\bibfnamefont {Y.}~\bibnamefont {Inoue}}, \
  and\ \bibinfo {author} {\bibfnamefont {M.~K.}\ \bibnamefont {Gilson}},\
  }\href {http://pubs.acs.org/doi/abs/10.1021/ja808175m
  papers://f7841d7d-9771-4011-86e7-7e86ad060f2b/Paper/p4484} {\bibfield
  {journal} {\bibinfo  {journal} {J Am Chem Soc}\
  }\textbf {\bibinfo {volume} {131}},\ \bibinfo {pages} {4012} (\bibinfo {year}
  {2009})}\BibitemShut {NoStop}%
\bibitem [{\citenamefont {Moghaddam}\ \emph {et~al.}(2011)\citenamefont
  {Moghaddam}, \citenamefont {Yang}, \citenamefont {Rekharsky}, \citenamefont
  {Ko}, \citenamefont {Kim}, \citenamefont {Inoue},\ and\ \citenamefont
  {Gilson}}]{Moghaddam2011}%
  \BibitemOpen
  \bibfield  {author} {\bibinfo {author} {\bibfnamefont {S.}~\bibnamefont
  {Moghaddam}}, \bibinfo {author} {\bibfnamefont {C.}~\bibnamefont {Yang}},
  \bibinfo {author} {\bibfnamefont {M.}~\bibnamefont {Rekharsky}}, \bibinfo
  {author} {\bibfnamefont {Y.~H.}\ \bibnamefont {Ko}}, \bibinfo {author}
  {\bibfnamefont {K.}~\bibnamefont {Kim}}, \bibinfo {author} {\bibfnamefont
  {Y.}~\bibnamefont {Inoue}}, \ and\ \bibinfo {author} {\bibfnamefont {M.~K.}\
  \bibnamefont {Gilson}},\ }\href
  {http://pubs.acs.org/doi/abs/10.1021/ja109904u
  papers://f7841d7d-9771-4011-86e7-7e86ad060f2b/Paper/p8255} {\bibfield
  {journal} {\bibinfo  {journal} {J Am Chem Soc}\
  }\textbf {\bibinfo {volume} {133}},\ \bibinfo {pages} {3570} (\bibinfo {year}
  {2011})}\BibitemShut {NoStop}%
\bibitem [{Note2()}]{Note2}%
  \BibitemOpen
  \bibinfo {note} {Using the linear combination of pairwise overlap \cite
  {Weiser1999} algorithm, NAMD 2.9 calculates a negative surface area for
  CB[7]. NAMD directly uses Appendix B of \protect \citet {Weiser1999}, in
  which the P$_1$ parameter for the N \protect \emph {sp3} atom type with 1
  bonded neighbor is $7.8602 \times 10^{-2}$, which is smaller than other P$_1$
  values and the corresponding P$_2$ parameter. By definition, P$_1$ should be
  larger than P$_2$. To bring this parameter in line with other P$_1$ and to
  make it larger than P$_2$, this parameter was multiplied by 10. The modified
  code yields a positive surface area for CB[7].}\BibitemShut {Stop}%
\bibitem [{\citenamefont {Phillips}\ \emph {et~al.}(2005)\citenamefont
  {Phillips}, \citenamefont {Braun}, \citenamefont {Wang}, \citenamefont
  {Gumbart}, \citenamefont {Tajkhorshid}, \citenamefont {Villa}, \citenamefont
  {Chipot}, \citenamefont {Skeel}, \citenamefont {Kal\'{e}},\ and\
  \citenamefont {Schulten}}]{Phillips2005}%
  \BibitemOpen
  \bibfield  {author} {\bibinfo {author} {\bibfnamefont {J.~C.}\ \bibnamefont
  {Phillips}}, \bibinfo {author} {\bibfnamefont {R.}~\bibnamefont {Braun}},
  \bibinfo {author} {\bibfnamefont {W.}~\bibnamefont {Wang}}, \bibinfo {author}
  {\bibfnamefont {J.}~\bibnamefont {Gumbart}}, \bibinfo {author} {\bibfnamefont
  {E.}~\bibnamefont {Tajkhorshid}}, \bibinfo {author} {\bibfnamefont
  {E.}~\bibnamefont {Villa}}, \bibinfo {author} {\bibfnamefont
  {C.}~\bibnamefont {Chipot}}, \bibinfo {author} {\bibfnamefont {R.~D.}\
  \bibnamefont {Skeel}}, \bibinfo {author} {\bibfnamefont {L.}~\bibnamefont
  {Kal\'{e}}}, \ and\ \bibinfo {author} {\bibfnamefont {K.}~\bibnamefont
  {Schulten}},\ }\href {\doibase 10.1002/jcc.20289} {\bibfield  {journal}
  {\bibinfo  {journal} {J Comput Chem}\ }\textbf {\bibinfo
  {volume} {26}},\ \bibinfo {pages} {1781} (\bibinfo {year}
  {2005})}\BibitemShut {NoStop}%
\bibitem [{\citenamefont {Davis}\ \emph {et~al.}(1991)\citenamefont {Davis},
  \citenamefont {Madura}, \citenamefont {Luty},\ and\ \citenamefont
  {McCammon}}]{Davis1991}%
  \BibitemOpen
  \bibfield  {author} {\bibinfo {author} {\bibfnamefont {M.~E.}\ \bibnamefont
  {Davis}}, \bibinfo {author} {\bibfnamefont {J.~D.}\ \bibnamefont {Madura}},
  \bibinfo {author} {\bibfnamefont {B.~a.}\ \bibnamefont {Luty}}, \ and\
  \bibinfo {author} {\bibfnamefont {J.}~\bibnamefont {McCammon}},\ }\href
  {\doibase 10.1016/0010-4655(91)90094-2} {\bibfield  {journal} {\bibinfo
  {journal} {Computer Physics Communications}\ }\textbf {\bibinfo {volume}
  {62}},\ \bibinfo {pages} {187} (\bibinfo {year} {1991})}\BibitemShut
  {NoStop}%
\bibitem [{\citenamefont {Duane}\ \emph {et~al.}(1987)\citenamefont {Duane},
  \citenamefont {Kennedy}, \citenamefont {Pendleton},\ and\ \citenamefont
  {Roweth}}]{DUANE1987}%
  \BibitemOpen
  \bibfield  {author} {\bibinfo {author} {\bibfnamefont {S.}~\bibnamefont
  {Duane}}, \bibinfo {author} {\bibfnamefont {A.~D.}\ \bibnamefont {Kennedy}},
  \bibinfo {author} {\bibfnamefont {B.~J.}\ \bibnamefont {Pendleton}}, \ and\
  \bibinfo {author} {\bibfnamefont {D.}~\bibnamefont {Roweth}},\ }\href@noop {}
  {\bibfield  {journal} {\bibinfo  {journal} {Phys Lett B}\ }\textbf
  {\bibinfo {volume} {195}},\ \bibinfo {pages} {216} (\bibinfo {year}
  {1987})}\BibitemShut {NoStop}%
\bibitem [{\citenamefont {Guvench}\ \emph {et~al.}(2001)\citenamefont
  {Guvench}, \citenamefont {Weiser}, \citenamefont {Shenkin}, \citenamefont
  {Kolossv�Ry},\ and\ \citenamefont {Still}}]{Guvench2001}%
  \BibitemOpen
  \bibfield  {author} {\bibinfo {author} {\bibfnamefont {O.}~\bibnamefont
  {Guvench}}, \bibinfo {author} {\bibfnamefont {J.}~\bibnamefont {Weiser}},
  \bibinfo {author} {\bibfnamefont {P.}~\bibnamefont {Shenkin}}, \bibinfo
  {author} {\bibfnamefont {I.}~\bibnamefont {Kolossv�Ry}}, \ and\ \bibinfo
  {author} {\bibfnamefont {W.}~\bibnamefont {Still}},\ }\href
  {papers://f7841d7d-9771-4011-86e7-7e86ad060f2b/Paper/p8040} {\bibfield
  {journal} {\bibinfo  {journal} {J Comput Chem}\ }\textbf
  {\bibinfo {volume} {23}},\ \bibinfo {pages} {214} (\bibinfo {year}
  {2001})}\BibitemShut {NoStop}%
\bibitem [{\citenamefont {Mobley}\ and\ \citenamefont
  {Dill}(2009)}]{Mobley2009}%
  \BibitemOpen
  \bibfield  {author} {\bibinfo {author} {\bibfnamefont {D.~L.}\ \bibnamefont
  {Mobley}}\ and\ \bibinfo {author} {\bibfnamefont {K.~A.}\ \bibnamefont
  {Dill}},\ }\href {http://dx.doi.org/10.1016/j.str.2009.02.010
  papers://f7841d7d-9771-4011-86e7-7e86ad060f2b/Paper/p4877} {\bibfield
  {journal} {\bibinfo  {journal} {Structure/Folding and Design}\ }\textbf
  {\bibinfo {volume} {17}},\ \bibinfo {pages} {489} (\bibinfo {year}
  {2009})}\BibitemShut {NoStop}%
\bibitem [{\citenamefont {Craig}\ \emph {et~al.}(2010)\citenamefont {Craig},
  \citenamefont {Essex},\ and\ \citenamefont {Spiegel}}]{Craig:2010p6676}%
  \BibitemOpen
  \bibfield  {author} {\bibinfo {author} {\bibfnamefont {I.~R.}\ \bibnamefont
  {Craig}}, \bibinfo {author} {\bibfnamefont {J.~W.}\ \bibnamefont {Essex}}, \
  and\ \bibinfo {author} {\bibfnamefont {K.}~\bibnamefont {Spiegel}},\ }\href
  {\doibase 10.1021/ci900407c} {\bibfield  {journal} {\bibinfo  {journal} {J
  Chem Inf Model}\ }\textbf {\bibinfo {volume} {50}},\ \bibinfo {pages} {511}
  (\bibinfo {year} {2010})}\BibitemShut {NoStop}%
\bibitem [{\citenamefont {Bottegoni}\ \emph {et~al.}(2011)\citenamefont
  {Bottegoni}, \citenamefont {Rocchia}, \citenamefont {Rueda}, \citenamefont
  {Abagyan},\ and\ \citenamefont {Cavalli}}]{Bottegoni:2011p6687}%
  \BibitemOpen
  \bibfield  {author} {\bibinfo {author} {\bibfnamefont {G.}~\bibnamefont
  {Bottegoni}}, \bibinfo {author} {\bibfnamefont {W.}~\bibnamefont {Rocchia}},
  \bibinfo {author} {\bibfnamefont {M.}~\bibnamefont {Rueda}}, \bibinfo
  {author} {\bibfnamefont {R.}~\bibnamefont {Abagyan}}, \ and\ \bibinfo
  {author} {\bibfnamefont {A.}~\bibnamefont {Cavalli}},\ }\href {\doibase
  10.1371/journal.pone.0018845.t005} {\bibfield  {journal} {\bibinfo  {journal}
  {PLoS ONE}\ }\textbf {\bibinfo {volume} {6}},\ \bibinfo {pages} {e18845}
  (\bibinfo {year} {2011})}\BibitemShut {NoStop}%
\bibitem [{\citenamefont {Nichols}\ \emph {et~al.}(2011)\citenamefont
  {Nichols}, \citenamefont {Baron}, \citenamefont {Ivetac},\ and\ \citenamefont
  {McCammon}}]{Nichols:2011p6682}%
  \BibitemOpen
  \bibfield  {author} {\bibinfo {author} {\bibfnamefont {S.~E.}\ \bibnamefont
  {Nichols}}, \bibinfo {author} {\bibfnamefont {R.}~\bibnamefont {Baron}},
  \bibinfo {author} {\bibfnamefont {A.}~\bibnamefont {Ivetac}}, \ and\ \bibinfo
  {author} {\bibfnamefont {J.~A.}\ \bibnamefont {McCammon}},\ }\href {\doibase
  10.1021/ci200117n} {\bibfield  {journal} {\bibinfo  {journal} {J Chem Inf
  Model}\ }\textbf {\bibinfo {volume} {51}},\ \bibinfo {pages} {1439} (\bibinfo
  {year} {2011})}\BibitemShut {NoStop}%
\bibitem [{\citenamefont {Lin}\ \emph {et~al.}(2002)\citenamefont {Lin},
  \citenamefont {Perryman}, \citenamefont {Schames},\ and\ \citenamefont
  {McCammon}}]{Lin:2002p6859}%
  \BibitemOpen
  \bibfield  {author} {\bibinfo {author} {\bibfnamefont {J.}~\bibnamefont
  {Lin}}, \bibinfo {author} {\bibfnamefont {A.}~\bibnamefont {Perryman}},
  \bibinfo {author} {\bibfnamefont {J.}~\bibnamefont {Schames}}, \ and\
  \bibinfo {author} {\bibfnamefont {J.}~\bibnamefont {McCammon}},\ }\href
  {\doibase 10.1021/ja0260162} {\bibfield  {journal} {\bibinfo  {journal} {J Am
  Chem Soc}\ }\textbf {\bibinfo {volume} {124}},\ \bibinfo {pages} {5632}
  (\bibinfo {year} {2002})}\BibitemShut {NoStop}%
\bibitem [{\citenamefont {Amaro}\ \emph {et~al.}(2008)\citenamefont {Amaro},
  \citenamefont {Baron},\ and\ \citenamefont {McCammon}}]{Amaro:2008p6685}%
  \BibitemOpen
  \bibfield  {author} {\bibinfo {author} {\bibfnamefont {R.~E.}\ \bibnamefont
  {Amaro}}, \bibinfo {author} {\bibfnamefont {R.}~\bibnamefont {Baron}}, \ and\
  \bibinfo {author} {\bibfnamefont {J.~A.}\ \bibnamefont {McCammon}},\ }\href
  {\doibase 10.1007/s10822-007-9159-2} {\bibfield  {journal} {\bibinfo
  {journal} {J Comput Aided Mol Des}\ }\textbf {\bibinfo {volume} {22}},\
  \bibinfo {pages} {693} (\bibinfo {year} {2008})}\BibitemShut {NoStop}%
\bibitem [{\citenamefont {Schames}\ \emph {et~al.}(2004)\citenamefont
  {Schames}, \citenamefont {Henchman}, \citenamefont {Siegel}, \citenamefont
  {Sotriffer}, \citenamefont {Ni},\ and\ \citenamefont
  {McCammon}}]{Schames:2004p6724}%
  \BibitemOpen
  \bibfield  {author} {\bibinfo {author} {\bibfnamefont {J.}~\bibnamefont
  {Schames}}, \bibinfo {author} {\bibfnamefont {R.}~\bibnamefont {Henchman}},
  \bibinfo {author} {\bibfnamefont {J.}~\bibnamefont {Siegel}}, \bibinfo
  {author} {\bibfnamefont {C.}~\bibnamefont {Sotriffer}}, \bibinfo {author}
  {\bibfnamefont {H.}~\bibnamefont {Ni}}, \ and\ \bibinfo {author}
  {\bibfnamefont {J.}~\bibnamefont {McCammon}},\ }\href@noop {} {\bibfield
  {journal} {\bibinfo  {journal} {J Med Chem}\ }\textbf {\bibinfo {volume}
  {47}},\ \bibinfo {pages} {1879} (\bibinfo {year} {2004})}\BibitemShut
  {NoStop}%
\bibitem [{\citenamefont {Amaro}\ \emph {et~al.}(2007)\citenamefont {Amaro},
  \citenamefont {Minh}, \citenamefont {Cheng}, \citenamefont {Lindstrom},
  \citenamefont {Olson}, \citenamefont {Lin}, \citenamefont {Li},\ and\
  \citenamefont {McCammon}}]{Amaro:2007p2}%
  \BibitemOpen
  \bibfield  {author} {\bibinfo {author} {\bibfnamefont {R.~E.}\ \bibnamefont
  {Amaro}}, \bibinfo {author} {\bibfnamefont {D.~D.~L.}\ \bibnamefont {Minh}},
  \bibinfo {author} {\bibfnamefont {L.~S.}\ \bibnamefont {Cheng}}, \bibinfo
  {author} {\bibfnamefont {W.~M.}\ \bibnamefont {Lindstrom}}, \bibinfo {author}
  {\bibfnamefont {A.~J.}\ \bibnamefont {Olson}}, \bibinfo {author}
  {\bibfnamefont {J.-H.}\ \bibnamefont {Lin}}, \bibinfo {author} {\bibfnamefont
  {W.~W.}\ \bibnamefont {Li}}, \ and\ \bibinfo {author} {\bibfnamefont {J.~A.}\
  \bibnamefont {McCammon}},\ }\href {\doibase 10.1021/ja0723535} {\bibfield
  {journal} {\bibinfo  {journal} {J Am Chem Soc}\ }\textbf {\bibinfo {volume}
  {129}},\ \bibinfo {pages} {7764} (\bibinfo {year} {2007})}\BibitemShut
  {NoStop}%
\bibitem [{\citenamefont {Durrant}\ and\ \citenamefont
  {McCammon}(2011)}]{Durrant:2011p6673}%
  \BibitemOpen
  \bibfield  {author} {\bibinfo {author} {\bibfnamefont {J.~D.}\ \bibnamefont
  {Durrant}}\ and\ \bibinfo {author} {\bibfnamefont {J.~A.}\ \bibnamefont
  {McCammon}},\ }\href {\doibase 10.1186/1741-7007-9-71} {\bibfield  {journal}
  {\bibinfo  {journal} {BMC Biol}\ }\textbf {\bibinfo {volume} {9}},\ \bibinfo
  {pages} {71} (\bibinfo {year} {2011})}\BibitemShut {NoStop}%
\bibitem [{\citenamefont {Paulsen}\ and\ \citenamefont
  {Anderson}(2009)}]{Paulsen2009}%
  \BibitemOpen
  \bibfield  {author} {\bibinfo {author} {\bibfnamefont {J.~L.}\ \bibnamefont
  {Paulsen}}\ and\ \bibinfo {author} {\bibfnamefont {A.~C.}\ \bibnamefont
  {Anderson}},\ }\href {\doibase 10.1021/ci9003078} {\bibfield  {journal}
  {\bibinfo  {journal} {Journal Chem Inf Model}\ }\textbf
  {\bibinfo {volume} {49}},\ \bibinfo {pages} {2813} (\bibinfo {year}
  {2009})}\BibitemShut {NoStop}%
\bibitem [{\citenamefont {Morris}\ \emph {et~al.}(1998)\citenamefont {Morris},
  \citenamefont {Goodsell}, \citenamefont {Halliday}, \citenamefont {Huey},
  \citenamefont {Hart}, \citenamefont {Belew},\ and\ \citenamefont
  {Olson}}]{Morris:1998p6664}%
  \BibitemOpen
  \bibfield  {author} {\bibinfo {author} {\bibfnamefont {G.}~\bibnamefont
  {Morris}}, \bibinfo {author} {\bibfnamefont {D.}~\bibnamefont {Goodsell}},
  \bibinfo {author} {\bibfnamefont {R.}~\bibnamefont {Halliday}}, \bibinfo
  {author} {\bibfnamefont {R.}~\bibnamefont {Huey}}, \bibinfo {author}
  {\bibfnamefont {W.}~\bibnamefont {Hart}}, \bibinfo {author} {\bibfnamefont
  {R.}~\bibnamefont {Belew}}, \ and\ \bibinfo {author} {\bibfnamefont
  {A.}~\bibnamefont {Olson}},\ }\href@noop {} {\bibfield  {journal} {\bibinfo
  {journal} {J Comput Chem}\ }\textbf {\bibinfo {volume} {19}},\ \bibinfo
  {pages} {1639} (\bibinfo {year} {1998})}\BibitemShut {NoStop}%
\bibitem [{\citenamefont {Kuntz}\ \emph {et~al.}(1982)\citenamefont {Kuntz},
  \citenamefont {Blaney}, \citenamefont {Oatley}, \citenamefont {Langridge},\
  and\ \citenamefont {Ferrin}}]{KUNTZ:1982p6867}%
  \BibitemOpen
  \bibfield  {author} {\bibinfo {author} {\bibfnamefont {I.~D.}\ \bibnamefont
  {Kuntz}}, \bibinfo {author} {\bibfnamefont {J.~M.}\ \bibnamefont {Blaney}},
  \bibinfo {author} {\bibfnamefont {S.~J.}\ \bibnamefont {Oatley}}, \bibinfo
  {author} {\bibfnamefont {R.}~\bibnamefont {Langridge}}, \ and\ \bibinfo
  {author} {\bibfnamefont {T.~E.}\ \bibnamefont {Ferrin}},\ }\href@noop {}
  {\bibfield  {journal} {\bibinfo  {journal} {J Mol Biol}\ }\textbf {\bibinfo
  {volume} {161}},\ \bibinfo {pages} {269} (\bibinfo {year}
  {1982})}\BibitemShut {NoStop}%
\bibitem [{\citenamefont {Goodsell}\ and\ \citenamefont
  {Olson}(1990)}]{Goodsell:1990p6732}%
  \BibitemOpen
  \bibfield  {author} {\bibinfo {author} {\bibfnamefont {D.}~\bibnamefont
  {Goodsell}}\ and\ \bibinfo {author} {\bibfnamefont {A.}~\bibnamefont
  {Olson}},\ }\href@noop {} {\bibfield  {journal} {\bibinfo  {journal}
  {Proteins}\ }\textbf {\bibinfo {volume} {8}},\ \bibinfo {pages} {195}
  (\bibinfo {year} {1990})}\BibitemShut {NoStop}%
\bibitem [{\citenamefont {Morris}\ \emph {et~al.}(1996)\citenamefont {Morris},
  \citenamefont {Goodsell}, \citenamefont {Huey},\ and\ \citenamefont
  {Olson}}]{Morris:1996p6733}%
  \BibitemOpen
  \bibfield  {author} {\bibinfo {author} {\bibfnamefont {G.}~\bibnamefont
  {Morris}}, \bibinfo {author} {\bibfnamefont {D.}~\bibnamefont {Goodsell}},
  \bibinfo {author} {\bibfnamefont {R.}~\bibnamefont {Huey}}, \ and\ \bibinfo
  {author} {\bibfnamefont {A.}~\bibnamefont {Olson}},\ }\href@noop {}
  {\bibfield  {journal} {\bibinfo  {journal} {J Comput Aided Mol Des}\ }\textbf
  {\bibinfo {volume} {10}},\ \bibinfo {pages} {293} (\bibinfo {year}
  {1996})}\BibitemShut {NoStop}%
\bibitem [{\citenamefont {Huey}\ \emph {et~al.}(2007)\citenamefont {Huey},
  \citenamefont {Morris}, \citenamefont {Olson},\ and\ \citenamefont
  {Goodsell}}]{Huey:2007p6666}%
  \BibitemOpen
  \bibfield  {author} {\bibinfo {author} {\bibfnamefont {R.}~\bibnamefont
  {Huey}}, \bibinfo {author} {\bibfnamefont {G.~M.}\ \bibnamefont {Morris}},
  \bibinfo {author} {\bibfnamefont {A.~J.}\ \bibnamefont {Olson}}, \ and\
  \bibinfo {author} {\bibfnamefont {D.~S.}\ \bibnamefont {Goodsell}},\ }\href
  {\doibase 10.1002/jcc.20634} {\bibfield  {journal} {\bibinfo  {journal} {J
  Comput Chem}\ }\textbf {\bibinfo {volume} {28}},\ \bibinfo {pages} {1145}
  (\bibinfo {year} {2007})}\BibitemShut {NoStop}%
\bibitem [{\citenamefont {Liu}\ and\ \citenamefont
  {Wang}(1999)}]{Liu:1999p6838}%
  \BibitemOpen
  \bibfield  {author} {\bibinfo {author} {\bibfnamefont {M.}~\bibnamefont
  {Liu}}\ and\ \bibinfo {author} {\bibfnamefont {S.}~\bibnamefont {Wang}},\
  }\href
  {http://apps.webofknowledge.com/InboundService.do?SID=3Dc4hd7abg3dJLHbd5E&product=WOS&UT=000081695300001&SrcApp=CR&DestFail=http%253A%252F%252Fwww.webofknowledge.com&Init=Yes&action=retrieve&Func=Frame&customersID=mekentosj&SrcAuth=mekentosj&IsProductCode=Yes&mode=FullRecord}
  {\bibfield  {journal} {\bibinfo  {journal} {J Comput Aided Mol Des}\ }\textbf
  {\bibinfo {volume} {13}},\ \bibinfo {pages} {435} (\bibinfo {year}
  {1999})}\BibitemShut {NoStop}%
\bibitem [{\citenamefont {Neal}(2001)}]{Neal:2001p3154}%
  \BibitemOpen
  \bibfield  {author} {\bibinfo {author} {\bibfnamefont {R.}~\bibnamefont
  {Neal}},\ }\href@noop {} {\bibfield  {journal} {\bibinfo  {journal} {Stat
  Comput}\ }\textbf {\bibinfo {volume} {11}},\ \bibinfo {pages} {125} (\bibinfo
  {year} {2001})}\BibitemShut {NoStop}%
\bibitem [{\citenamefont {Carlson}(2002)}]{Carlson:2002p6715}%
  \BibitemOpen
  \bibfield  {author} {\bibinfo {author} {\bibfnamefont {H.}~\bibnamefont
  {Carlson}},\ }\href@noop {} {\bibfield  {journal} {\bibinfo  {journal} {Curr
  Opin Chem Biol}\ }\textbf {\bibinfo {volume} {6}},\ \bibinfo {pages} {447}
  (\bibinfo {year} {2002})}\BibitemShut {NoStop}%
\bibitem [{\citenamefont {Xu}\ \emph {et~al.}(2008)\citenamefont {Xu},
  \citenamefont {Colletier}, \citenamefont {Jiang}, \citenamefont {Silman},
  \citenamefont {Sussman},\ and\ \citenamefont {Weik}}]{Xu:2008p6806}%
  \BibitemOpen
  \bibfield  {author} {\bibinfo {author} {\bibfnamefont {Y.}~\bibnamefont
  {Xu}}, \bibinfo {author} {\bibfnamefont {J.~P.}\ \bibnamefont {Colletier}},
  \bibinfo {author} {\bibfnamefont {H.}~\bibnamefont {Jiang}}, \bibinfo
  {author} {\bibfnamefont {I.}~\bibnamefont {Silman}}, \bibinfo {author}
  {\bibfnamefont {J.~L.}\ \bibnamefont {Sussman}}, \ and\ \bibinfo {author}
  {\bibfnamefont {M.}~\bibnamefont {Weik}},\ }\href {\doibase
  10.1110/ps.083453808} {\bibfield  {journal} {\bibinfo  {journal} {Protein
  Sci}\ }\textbf {\bibinfo {volume} {17}},\ \bibinfo {pages} {601} (\bibinfo
  {year} {2008})}\BibitemShut {NoStop}%
\bibitem [{\citenamefont {Bucher}\ \emph {et~al.}(2011)\citenamefont {Bucher},
  \citenamefont {Grant},\ and\ \citenamefont {McCammon}}]{Bucher:2011p6810}%
  \BibitemOpen
  \bibfield  {author} {\bibinfo {author} {\bibfnamefont {D.}~\bibnamefont
  {Bucher}}, \bibinfo {author} {\bibfnamefont {B.~J.}\ \bibnamefont {Grant}}, \
  and\ \bibinfo {author} {\bibfnamefont {J.~A.}\ \bibnamefont {McCammon}},\
  }\href {\doibase 10.1021/bi201481a} {\bibfield  {journal} {\bibinfo
  {journal} {Biochemistry-Us}\ }\textbf {\bibinfo {volume} {50}},\ \bibinfo
  {pages} {10530} (\bibinfo {year} {2011})}\BibitemShut {NoStop}%
\bibitem [{\citenamefont {Weiser}\ \emph {et~al.}(1999)\citenamefont {Weiser},
  \citenamefont {Shenkin},\ and\ \citenamefont {Still}}]{Weiser1999}%
  \BibitemOpen
  \bibfield  {author} {\bibinfo {author} {\bibfnamefont {J.}~\bibnamefont
  {Weiser}}, \bibinfo {author} {\bibfnamefont {P.~S.}\ \bibnamefont {Shenkin}},
  \ and\ \bibinfo {author} {\bibfnamefont {W.~C.}\ \bibnamefont {Still}},\
  }\href {\doibase
  10.1002/(SICI)1096-987X(19990130)20:2<217::AID-JCC4>3.0.CO;2-A} {\bibfield
  {journal} {\bibinfo  {journal} {J Comput Chem}\ }\textbf
  {\bibinfo {volume} {20}},\ \bibinfo {pages} {217} (\bibinfo {year}
  {1999})}\BibitemShut {NoStop}%
\bibitem [{\citenamefont {Wagoner}\ and\ \citenamefont
  {Pande}(2011)}]{Wagoner2011}%
  \BibitemOpen
  \bibfield  {author} {\bibinfo {author} {\bibfnamefont {J.~A.}\ \bibnamefont
  {Wagoner}}\ and\ \bibinfo {author} {\bibfnamefont {V.~S.}\ \bibnamefont
  {Pande}},\ }\href {papers://f7841d7d-9771-4011-86e7-7e86ad060f2b/Paper/p6634}
  {\bibfield  {journal} {\bibinfo  {journal} {J Chem Phys}\
  }\textbf {\bibinfo {volume} {134}},\ \bibinfo {pages} {214103} (\bibinfo
  {year} {2011})}\BibitemShut {NoStop}%
\bibitem [{\citenamefont {Wong}\ \emph {et~al.}(2009)\citenamefont {Wong},
  \citenamefont {Amaro},\ and\ \citenamefont {McCammon}}]{Wong:2009p6414}%
  \BibitemOpen
  \bibfield  {author} {\bibinfo {author} {\bibfnamefont {S.}~\bibnamefont
  {Wong}}, \bibinfo {author} {\bibfnamefont {R.~E.}\ \bibnamefont {Amaro}}, \
  and\ \bibinfo {author} {\bibfnamefont {J.~A.}\ \bibnamefont {McCammon}},\
  }\href {\doibase 10.1021/ct8003707} {\bibfield  {journal} {\bibinfo
  {journal} {J Chem Theory Comput}\ }\textbf {\bibinfo {volume} {5}},\ \bibinfo
  {pages} {422} (\bibinfo {year} {2009})}\BibitemShut {NoStop}%
\end{thebibliography}
%

\newpage

\section{Supplemental Material}

\setcounter{figure}{0}
\makeatletter 
\renewcommand{\thefigure}{S\@arabic\c@figure} 
\makeatother

\setcounter{table}{0}
\makeatletter 
\renewcommand{\thetable}{S\@Roman\c@table} 
\makeatother

This supplemental material contains one theoretical section, two tables, and three figures.  The theoretical section describes a hybrid implicit-explicit solvent model.  One table is for components of the binding PMF and the other for average potential energies.  The figures show the convergence of binding PMF and binding free energy estimates, as well histogram of binding PMF estimates for different receptor snapshots.

\subsection{Hybrid Implicit-Explicit Solvent}

The desire to combine the speed of implicit solvent with the molecular detail and accuracy of explicit solvent has inspired interest in hybrid implicit-explicit solvent models (see \citet{Wagoner2011} and references therein).
In the context of implicit ligand theory, a small number of explicit solvent molecules can be considered as a part of the receptor \cite{Wong:2009p6414} during binding PMF calculations.

A simple formalism for a hybrid implicit-explicit solvent model may be derived by separating the coordinates of $N$ solvent molecules $r_S$ into explicitly represented coordinates $r_E$ and implicitly represented coordinates $r_I$.  Partition functions analogous to and formally equivalent to Eqs. (\ref{eq:Z_RL}) and (\ref{eq:Z_Y}) are then defined as,
\begin{eqnarray}
Z_{RL'} & = & \int  I_\delta e^{-\beta [U(r_{RL},r_E) + W(r_{RL},r_E)]} dr_{RL} dr_E
\label{eq:Z_RL'}\\
Z_{R'} & = & \int e^{-\beta [U(r_R,r_E) + W(r_R,r_E)]} dr_R dr_E.
\label{eq:Z_Y'}
\end{eqnarray}
Defining the effective interaction energy as $\Psi'(r_{RL},r_E) = \U(r_{RL},r_E) - \U(r_R,r_E) - \U(r_L)$ and the binding PMF as,
\begin{eqnarray}
B'(r_R,r_E) & = & -\beta^{-1} \ln \left( \frac{\int I_\delta e^{-\beta [\Psi(r_{RL},r_E) + \U(r_L)]} dr_L d\delta_L}{ \int I_\delta e^{-\beta \U(r_L)} dr_L d\delta_L } \right) \nonumber,
\end{eqnarray}
the binding free energy may be written as,
\begin{eqnarray}
\Delta G^\circ 
&=& -\beta^{-1} \ln \left( \frac{Z_{RL'}}{Z_R' Z_L} \frac{C^\circ}{8 \pi^2} \right) \nonumber \\
&=& -\beta^{-1} \ln \left( \frac{\int e^{-\beta[B'(r_R,r_E)+\U(r_R,r_E)]} dr_R dr_E}{ \int e^{-\beta \U(r_R,r_E)} dr_R dr_E } \frac{\Omega C^\circ}{8 \pi^2} \right) \nonumber \\
&=& -\beta^{-1} \ln \left< e^{-\beta B'} \right>^{r_R,r_E}_{R,E} + \Delta G_\delta,
\end{eqnarray}
where $q_{R,E} = e^{-\beta \U(r_R,r_E)}$.  The main text focuses on describing calculations in implicit solvent, with the understanding that explicit solvent may be readily included.

\newpage

\begin{sidewaystable}[h]
\center
\begin{tabular}{ l | c | c | c c | c c | c c | c c }
Ligand & Charge\footnote{Net formal charge} & $B_{cpl}$& $B_{RL,NAMD}$& $B_{L,NAMD}$& $B_{RL,M2}$& $B_{L,M2}$& $B_{RL,PB}$& $B_{L,PB}$& $B_{RL,PBSA}$& $B_{L,PBSA}$ \\ 
\hline
AD1 & 0 & -31.6 (0.09) & -115.9 (0.78) & -2.8 (0.01) & -111.5 (0.50) & -1.1 (0.01) & -128.1 (0.81) & -4.5 (0.01) & -122.3 (0.81) & -2.6 (0.01) \\ 
AD2 & 1 & -91.3 (0.12) & -123.7 (0.10) & -51.9 (0.04) & -112.8 (0.04) & -55.2 (0.04) & -123.4 (0.02) & -55.8 (0.03) & -117.6 (0.02) & -53.8 (0.03) \\ 
AD3 & 1 & -93.2 (0.17) & -118.9 (0.06) & -50.5 (0.04) & -108.9 (0.05) & -51.4 (0.04) & -122.3 (0.10) & -54.5 (0.04) & -116.4 (0.10) & -52.3 (0.04) \\ 
AD4 & 2 & -148.0 (0.41) & -202.5 (0.62) & -175.9 (0.63) & -192.2 (0.75) & -183.5 (0.58) & -188.8 (0.87) & -180.6 (0.54) & -182.8 (0.86) & -178.1 (0.54) \\ 
AD5 & 1 & -90.5 (0.13) & -124.3 (0.76) & -52.1 (0.05) & -111.9 (0.29) & -53.5 (0.02) & -123.5 (0.03) & -55.9 (0.03) & -117.6 (0.03) & -53.9 (0.03) \\ 
B02 & 0 & -31.9 (0.09) & -118.4 (0.37) & -6.9 (0.05) & -111.4 (0.33) & -4.5 (0.03) & -129.1 (0.49) & -9.0 (0.02) & -123.3 (0.49) & -6.8 (0.02) \\ 
B05 & 2 & -156.8 (0.13) & -187.3 (0.27) & -173.2 (0.11) & -176.4 (0.35) & -182.7 (0.18) & -173.4 (0.45) & -178.6 (0.15) & -167.5 (0.45) & -176.4 (0.15) \\ 
B11 & 4 & -220.7 (0.83) & -437.9 (1.24) & -475.7 (0.65) & -422.5 (1.20) & -484.7 (0.97) & -404.2 (1.93) & -478.7 (0.74) & -396.9 (1.80) & -474.3 (0.75) \\ 
F01 & 0 & -25.2 (0.27) & -117.2 (1.50) & -10.1 (0.10) & -65.0 (0.75) & 35.3 (0.07) & -92.5 (0.45) & 25.4 (0.07) & -86.6 (0.46) & 27.6 (0.08) \\ 
F02 & 1 & -84.2 (0.24) & -112.1 (0.87) & -50.9 (0.07) & -77.9 (0.57) & -28.0 (0.11) & -96.6 (0.32) & -32.5 (0.08) & -90.7 (0.32) & -29.9 (0.08) \\ 
F03 & 1 & -82.1 (0.51) & -111.8 (2.04) & -47.0 (0.04) & -76.2 (1.45) & -25.2 (0.07) & -97.0 (1.37) & -30.5 (0.05) & -91.1 (1.36) & -27.9 (0.05) \\ 
F06 & 2 & -144.6 (0.15) & -152.3 (0.09) & -135.7 (0.10) & -123.3 (0.12) & -127.9 (0.18) & -139.0 (0.09) & -129.5 (0.14) & -132.9 (0.09) & -126.3 (0.14) \\ 
\end{tabular}
\caption{The mean and standard deviation of 15 independent estimates for components of the binding PMF $B(r_R)$ (kcal/mol), as described in Eq. (\ref{eq:bindingPMF_decomposition}), for various ligands to the minimized structure of Cu[7].
\label{tab:S1}}
\end{sidewaystable}

\begin{table}
\center
Average Potential Energy of Complexes\\
\begin{tabular}{l | c c c c c c }
Ligand & VDW & Coul & PB & Val & NP & Total \\ 
\hline
AD1 & -90.2 (0.417) &  50.7 (1.415) & -131.6 (1.500) & 328.2 (2.651) &   5.8 (0.011) & 162.8 (2.400) \\ 
AD2 & -91.9 (0.905) &  29.7 (0.888) & -122.9 (0.687) & 328.5 (1.710) &   5.8 (0.017) & 149.2 (1.464) \\ 
AD3 & -91.7 (0.684) &  23.7 (0.452) & -124.3 (0.768) & 333.8 (1.950) &   5.8 (0.009) & 147.4 (2.231) \\ 
AD4 & -98.1 (1.382) &  79.7 (3.240) & -191.4 (0.930) & 348.1 (4.394) &   6.1 (0.070) & 144.4 (2.685) \\ 
AD5 & -91.5 (1.357) &  19.7 (1.457) & -123.1 (1.364) & 331.6 (1.565) &   5.8 (0.025) & 142.5 (1.624) \\ 
B02 & -87.2 (0.582) &  39.8 (0.929) & -131.2 (0.672) & 340.6 (3.270) &   5.8 (0.021) & 167.8 (2.014) \\ 
B05 & -90.5 (0.869) &  29.4 (1.113) & -173.0 (1.087) & 340.4 (0.896) &   5.8 (0.012) & 112.1 (1.440) \\ 
B11 & -103.0 (1.171) & 336.9 (2.218) & -404.0 (0.997) & 379.8 (5.364) &   7.3 (0.075) & 217.0 (4.393) \\ 
F01 & -88.1 (0.446) &  84.0 (1.747) & -127.1 (1.054) & 515.7 (4.767) &   5.8 (0.016) & 390.4 (3.989) \\ 
F02 & -89.4 (1.502) &  56.6 (2.010) & -119.3 (0.843) & 514.2 (2.068) &   5.9 (0.012) & 368.1 (2.086) \\ 
F03 & -90.1 (0.802) &  55.2 (0.835) & -117.4 (0.529) & 518.1 (3.448) &   5.9 (0.015) & 371.7 (3.320) \\ 
F06 & -97.3 (1.133) &  63.0 (0.616) & -153.2 (0.532) & 525.4 (4.787) &   6.1 (0.012) & 343.9 (4.044) \\ 
\end{tabular}
\\ \bigskip Average Potential Energy of Ligands\\
\begin{tabular}{l | c c c c c c }
Ligand & VDW & Coul & PB & Val & NP & Total \\ 
\hline
AD1 &  -1.8 (0.0063) &  -8.5 (0.0012) &  -4.6 (0.0014) &  52.6 (0.0232) &   2.0 (0.0001) &  39.7 (0.0236) \\ 
AD2 &  -2.5 (0.0049) &  36.3 (0.0036) & -55.9 (0.0028) &  53.8 (0.0246) &   2.0 (0.0001) &  33.7 (0.0248) \\ 
AD3 &  -3.0 (0.0058) &  29.0 (0.0028) & -54.6 (0.0028) &  58.8 (0.0259) &   2.1 (0.0001) &  32.4 (0.0264) \\ 
AD4 &  -4.1 (0.0056) & 145.8 (0.0124) & -183.9 (0.0089) &  65.6 (0.0291) &   2.5 (0.0001) &  25.9 (0.0270) \\ 
AD5 &  -2.4 (0.0052) &  25.6 (0.0032) & -56.1 (0.0030) &  55.9 (0.0245) &   2.0 (0.0001) &  25.0 (0.0248) \\ 
B02 &   1.9 (0.0103) & -13.6 (0.0032) &  -9.1 (0.0025) &  58.7 (0.0257) &   2.2 (0.0001) &  40.1 (0.0255) \\ 
B05 &  -1.7 (0.0069) & 108.3 (0.0081) & -179.2 (0.0055) &  62.1 (0.0270) &   2.3 (0.0001) &  -8.2 (0.0273) \\ 
B11 &  -7.2 (0.0116) & 477.0 (0.0132) & -484.1 (0.0106) & 104.8 (0.0268) &   4.4 (0.0002) &  94.8 (0.0232) \\ 
F01 &  -6.0 (0.0036) &  33.1 (0.0059) &  -9.5 (0.0036) & 226.5 (0.0251) &   2.2 (0.0001) & 246.4 (0.0258) \\ 
F02 &  -6.5 (0.0055) &  63.1 (0.0069) & -53.4 (0.0037) & 234.6 (0.0268) &   2.6 (0.0002) & 240.4 (0.0270) \\ 
F03 &  -5.5 (0.0078) &  54.1 (0.0069) & -49.8 (0.0035) & 238.0 (0.0286) &   2.6 (0.0001) & 239.4 (0.0284) \\ 
F06 &  -6.4 (0.0101) & 120.0 (0.0109) & -142.3 (0.0064) & 253.6 (0.0341) &   3.3 (0.0002) & 228.1 (0.0339) \\ 
\end{tabular}
\\ \bigskip Average Potential Energy of the Receptor\\
\begin{tabular}{c c c c c c }
VDW & Coul & PB & Val & NP & Total \\ 
\hline
 -55.9 (0.2198) &  59.2 (0.5253) & -131.8 (0.3756) & 280.6 (0.8513) &   6.3 (0.0039) & 158.4 (0.8516) \\ 
\end{tabular}
\\ \bigskip Average Potential Energy Changes\\
\begin{tabular}{l | c c c c c c }
Ligand & VDW & Coul & PB & Val & NP & Total \\ 
\hline
AD1 & -32.5 (0.471) &   0.1 (1.509) &   4.8 (1.547) &  -5.0 (2.785) &  -2.5 (0.011) & -35.2 (2.547) \\ 
AD2 & -33.6 (0.931) & -65.8 (1.032) &  64.9 (0.783) &  -5.9 (1.910) &  -2.5 (0.017) & -42.9 (1.693) \\ 
AD3 & -32.8 (0.718) & -64.4 (0.693) &  62.2 (0.855) &  -5.7 (2.128) &  -2.6 (0.009) & -43.4 (2.388) \\ 
AD4 & -38.1 (1.400) & -125.2 (3.283) & 124.4 (1.003) &   1.9 (4.475) &  -2.7 (0.070) & -39.9 (2.817) \\ 
AD5 & -33.3 (1.374) & -65.1 (1.549) &  64.8 (1.415) &  -4.9 (1.782) &  -2.5 (0.025) & -40.9 (1.834) \\ 
B02 & -33.3 (0.622) &  -5.8 (1.067) &   9.7 (0.770) &   1.4 (3.379) &  -2.7 (0.022) & -30.6 (2.187) \\ 
B05 & -32.9 (0.896) & -138.2 (1.231) & 138.0 (1.151) &  -2.3 (1.236) &  -2.8 (0.013) & -38.1 (1.673) \\ 
B11 & -39.9 (1.192) & -199.3 (2.280) & 212.0 (1.066) &  -5.6 (5.431) &  -3.4 (0.075) & -36.2 (4.475) \\ 
F01 & -26.2 (0.497) &  -8.2 (1.824) &  14.2 (1.119) &   8.7 (4.842) &  -2.7 (0.017) & -14.3 (4.079) \\ 
F02 & -26.9 (1.518) & -65.7 (2.078) &  65.9 (0.923) &  -0.9 (2.237) &  -3.0 (0.012) & -30.6 (2.253) \\ 
F03 & -28.7 (0.832) & -58.0 (0.987) &  64.2 (0.649) &  -0.4 (3.552) &  -3.0 (0.015) & -26.1 (3.428) \\ 
F06 & -35.1 (1.154) & -116.1 (0.810) & 120.9 (0.651) &  -8.8 (4.862) &  -3.5 (0.013) & -42.6 (4.132) \\ 
\end{tabular}
\caption{
Estimates of the mean potential energy (kcal/mol) of the ligand, receptor, and complex for different Cu[7] ligands.
The columns refer to van der Waals (VDW), coulomb (Coul), electrostatic solvation (PB), valence (Val, bond + angle + dihedral), nonpolar solvation (NP), and total energies.
The value in the parentheses is the standard deviation from bootstrapping: the observable is estimated based on 1000 random selections of 100 values of $\hat{\Theta}$.
\label{tab:S2}}
\end{table}

\newpage 

\begin{figure}
\subfloat{\includegraphics{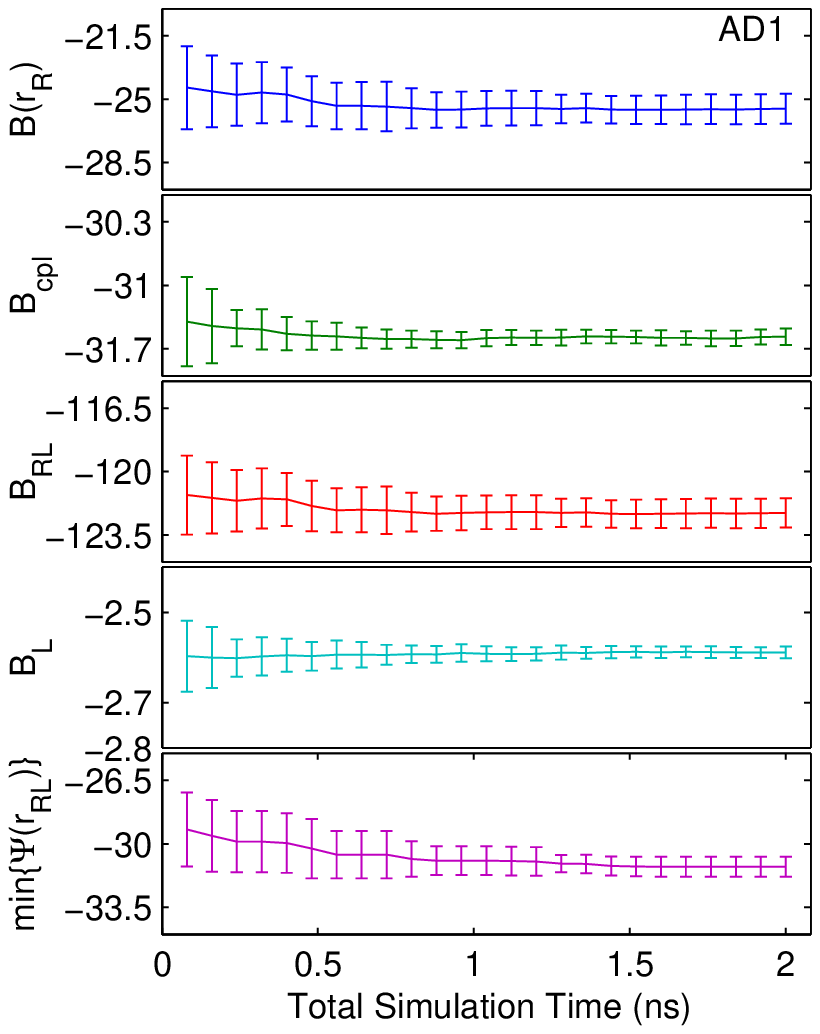}}
\subfloat{\includegraphics{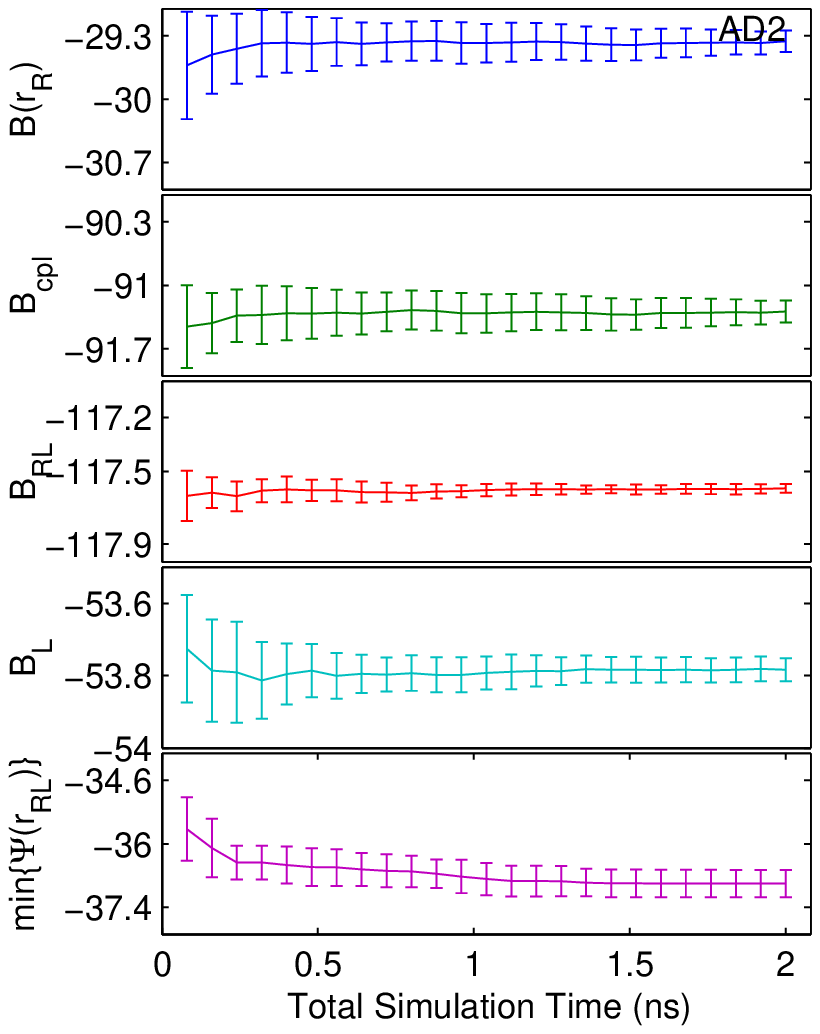}} \,
\subfloat{\includegraphics{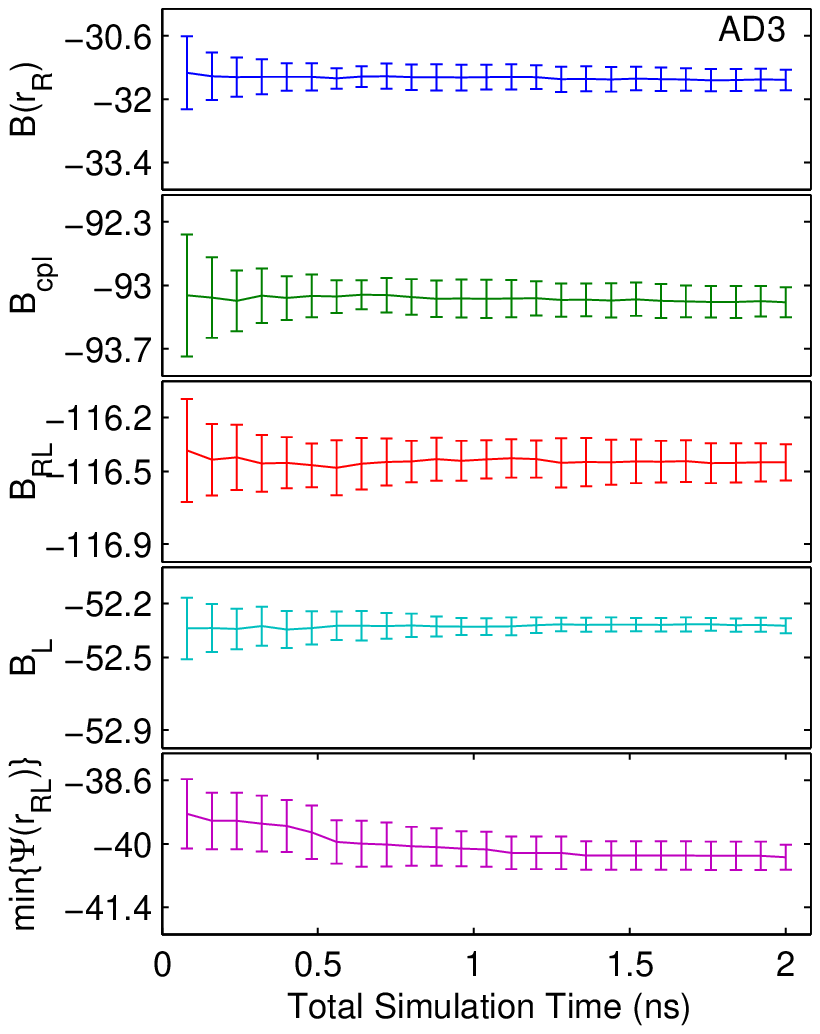}}
\subfloat{\includegraphics{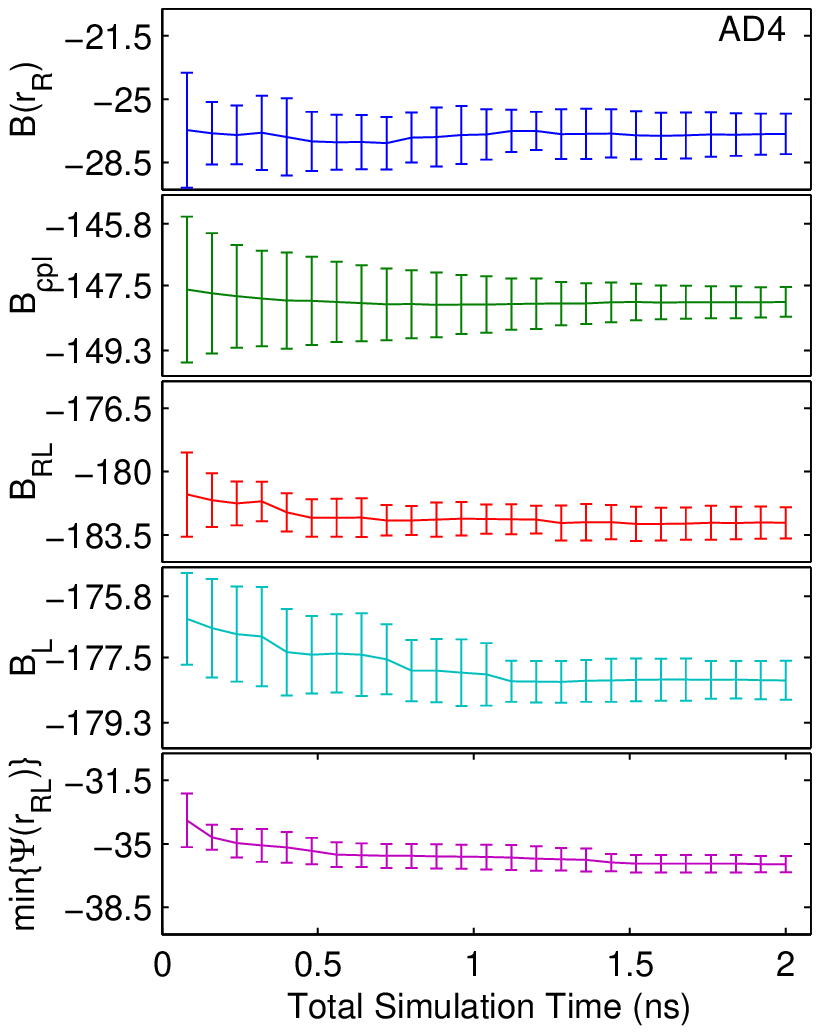}}
\caption{(a) The mean and standard deviation of 15 independent estimates of $B(r_R)$, $B_{cpl}$, $B_{RL}$, $B_L$, and min$\left\{\Psi(r_{RL})\right\}$ (kcal/mol) based on PBSA energies as a function of total MD simulation time.
\label{fig:S1}}
\end{figure}

\begin{figure}
\ContinuedFloat
\subfloat{\includegraphics{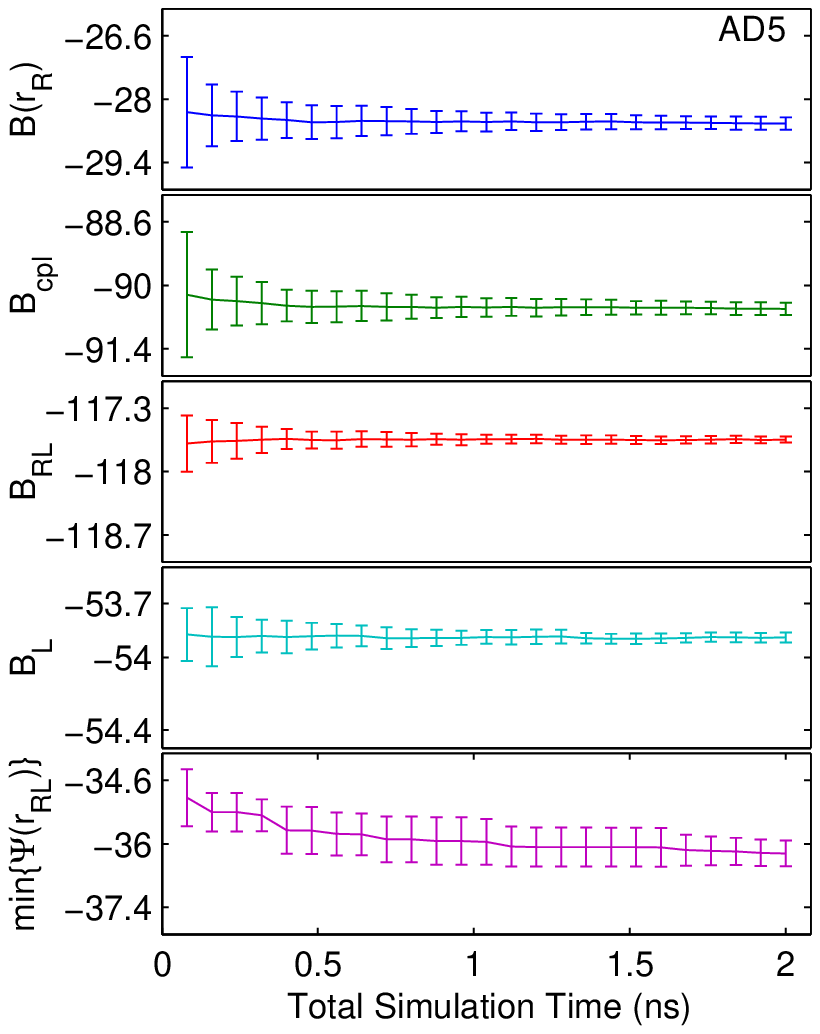}}
\subfloat{\includegraphics{fig_Bconvergence_f.eps}} \,
\subfloat{\includegraphics{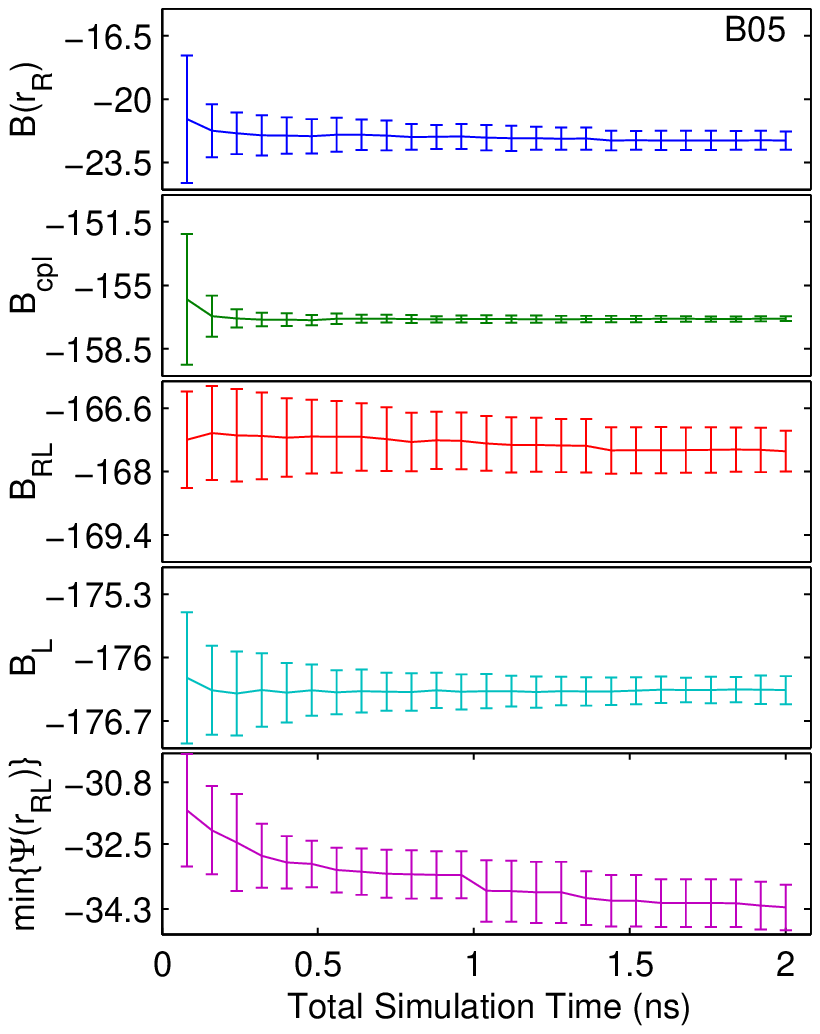}}
\subfloat{\includegraphics{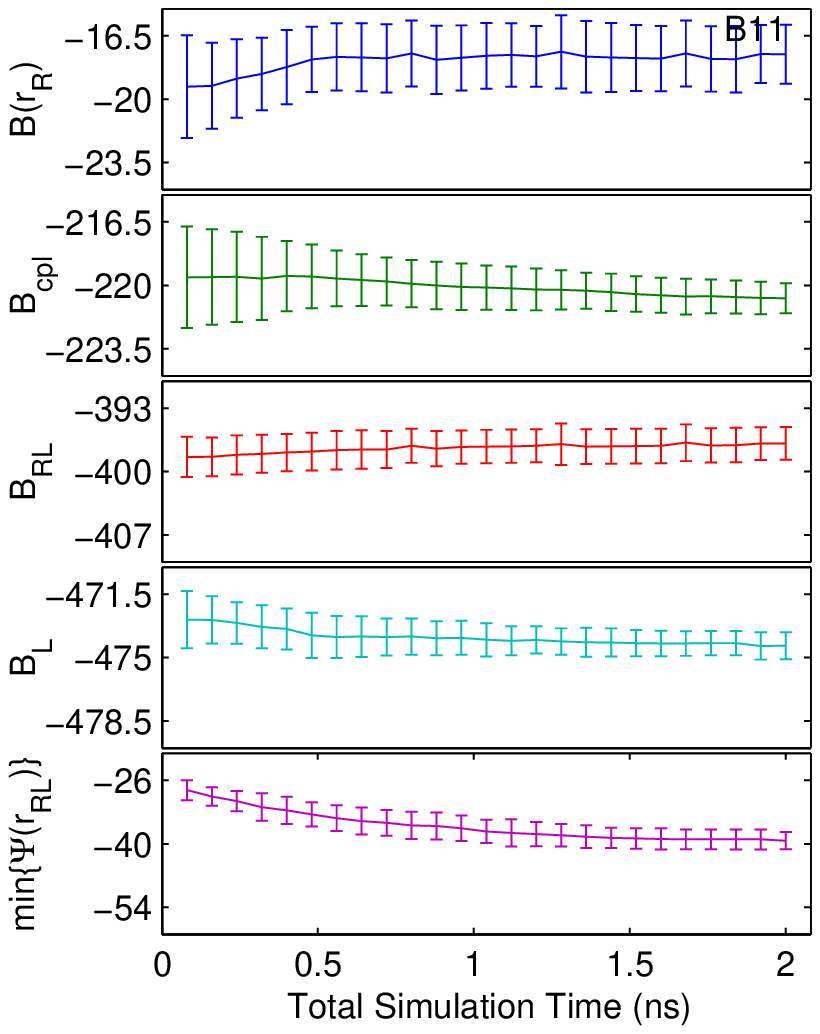}}
\caption{(b) The mean and standard deviation of 15 independent estimates of $B(r_R)$, $B_{cpl}$, $B_{RL}$, $B_L$, and min$\left\{\Psi(r_{RL})\right\}$ (kcal/mol) based on PBSA energies as a function of total MD simulation time.}
\end{figure}

\begin{figure}
\ContinuedFloat
\subfloat{\includegraphics{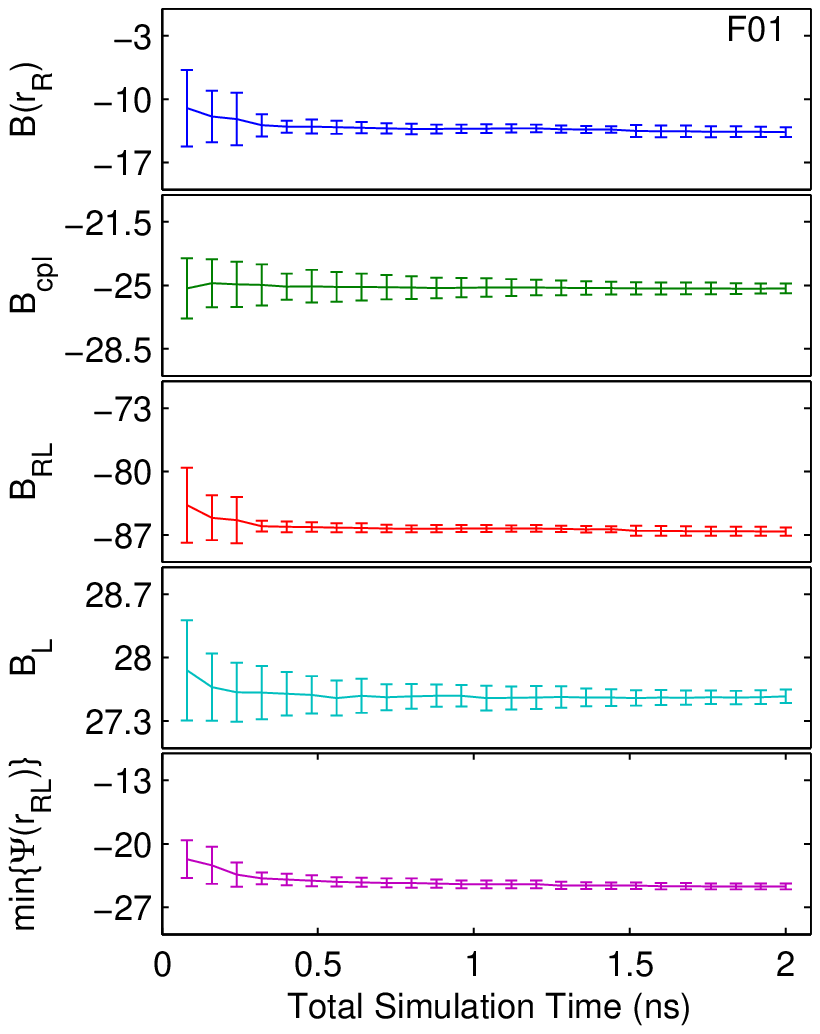}}
\subfloat{\includegraphics{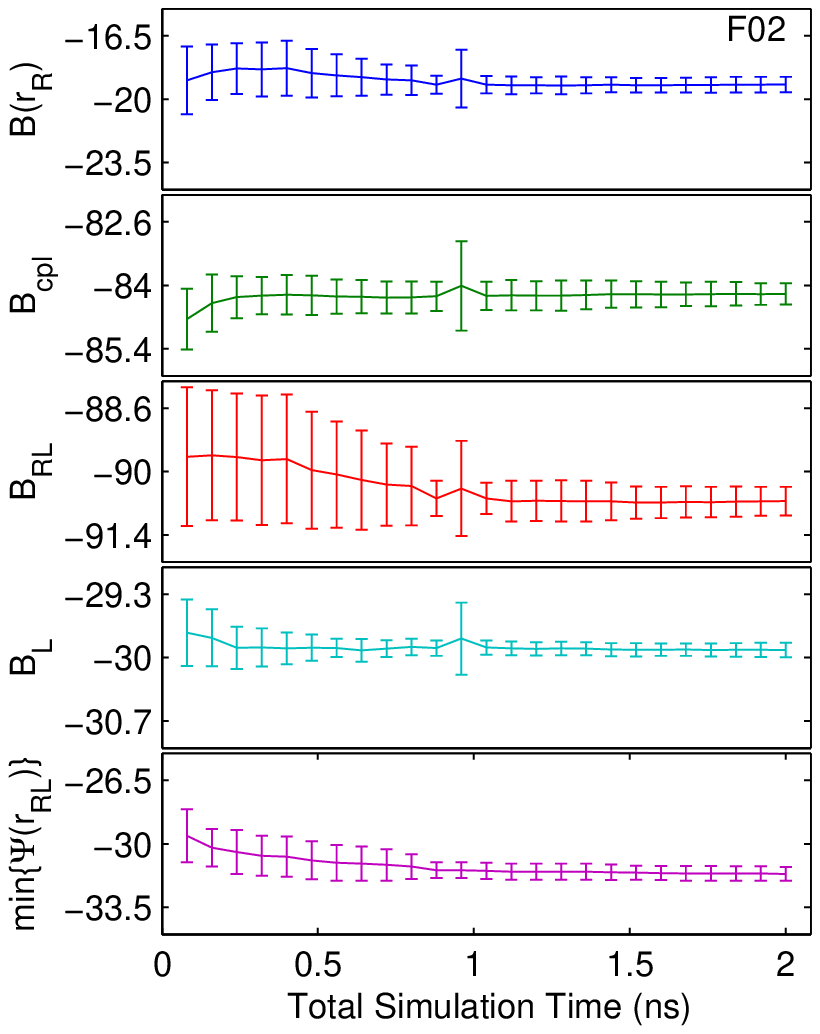}} \,
\subfloat{\includegraphics{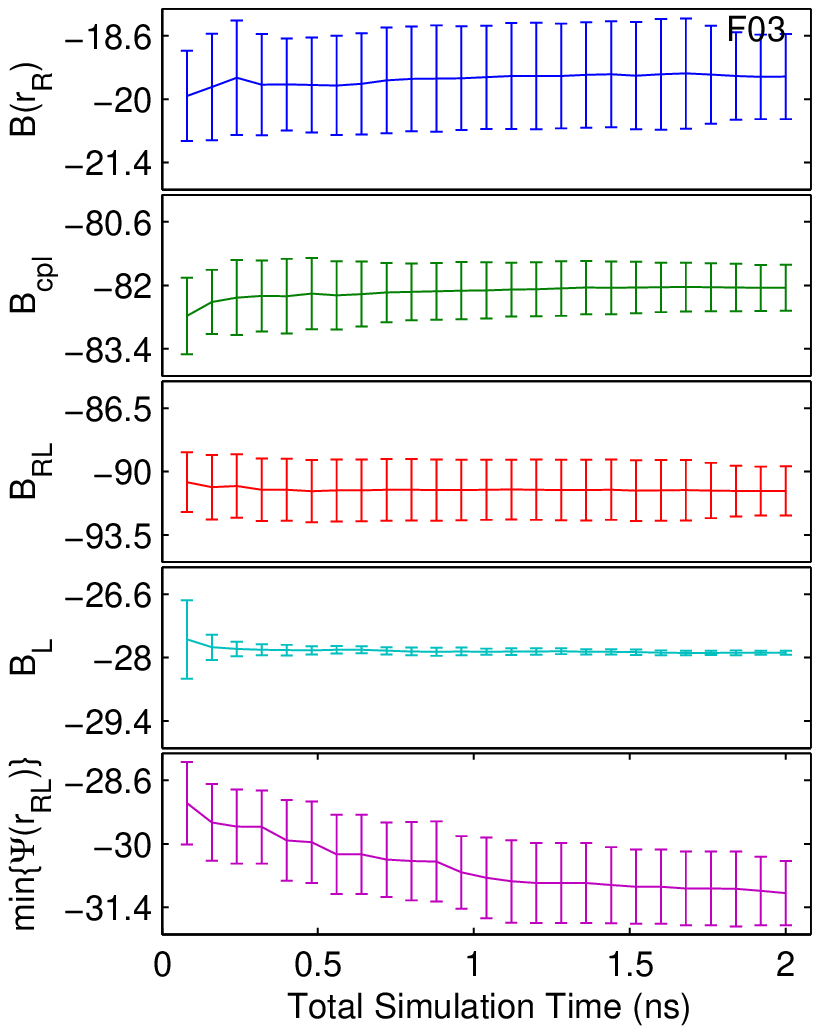}}
\subfloat{\includegraphics{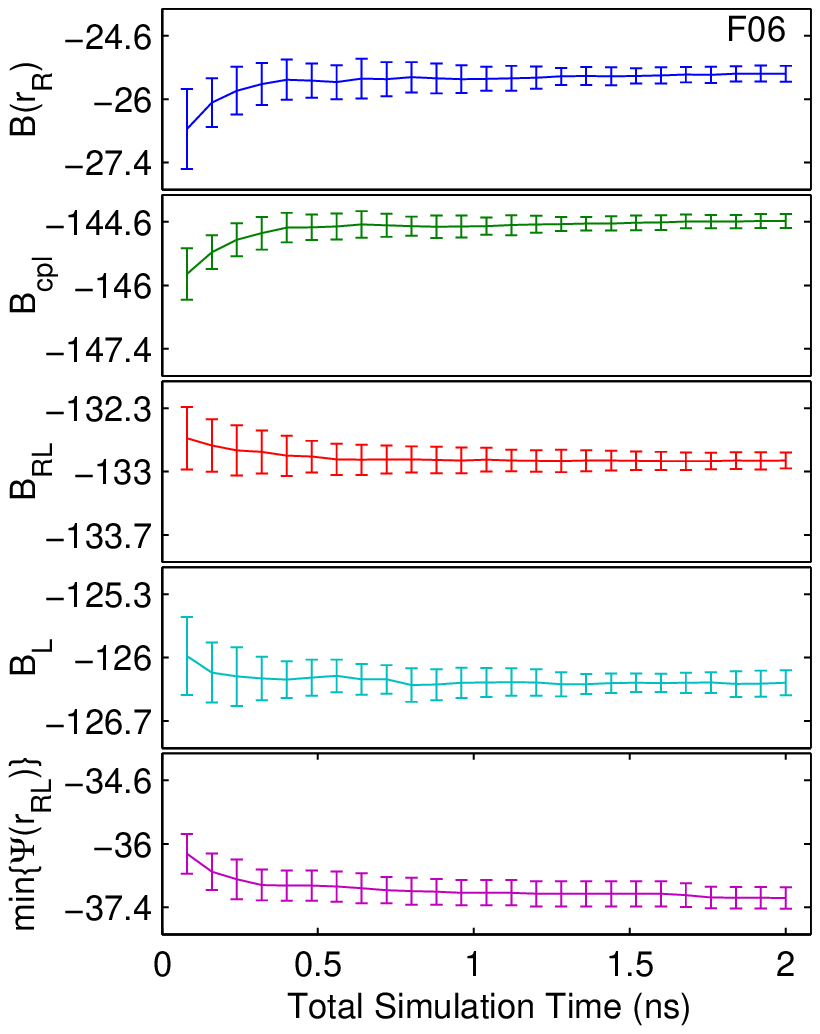}}
\caption{(c) The mean and standard deviation of 15 independent estimates of $B(r_R)$, $B_{cpl}$, $B_{RL}$, $B_L$, and min$\left\{\Psi(r_{RL})\right\}$ (kcal/mol) based on PBSA energies as a function of total MD simulation time.}
\end{figure}

\begin{figure}
\includegraphics{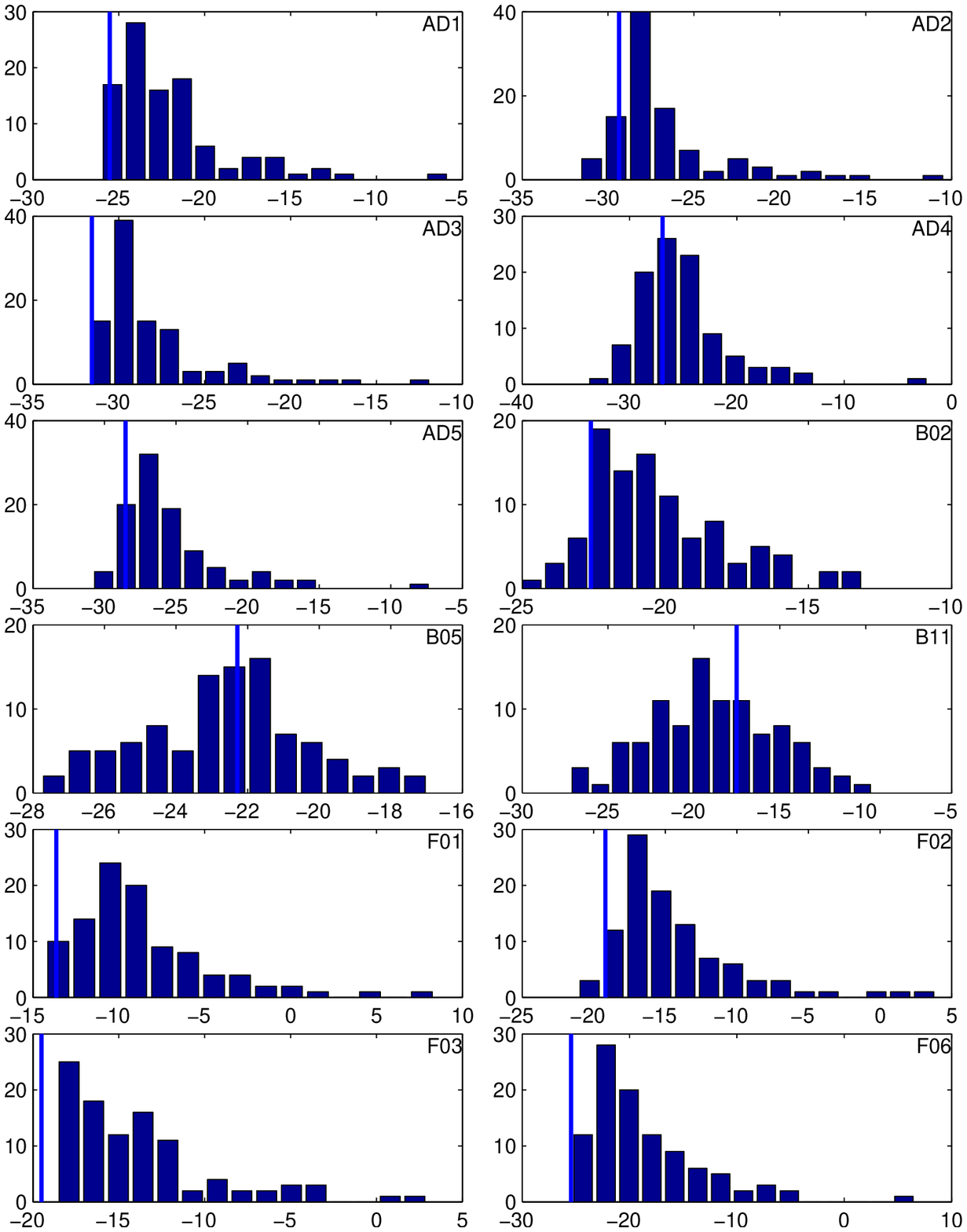}
\caption{
Histogram of binding PMF estimates $\hat{B}(r_R)$ (kcal/mol) of various ligands to 100 snapshots of Cu[7], using PBSA energies.  The vertical line shows the mean binding PMF for the minimized receptor structure.
\label{fig:S2}}
\end{figure}

\begin{figure}
\includegraphics{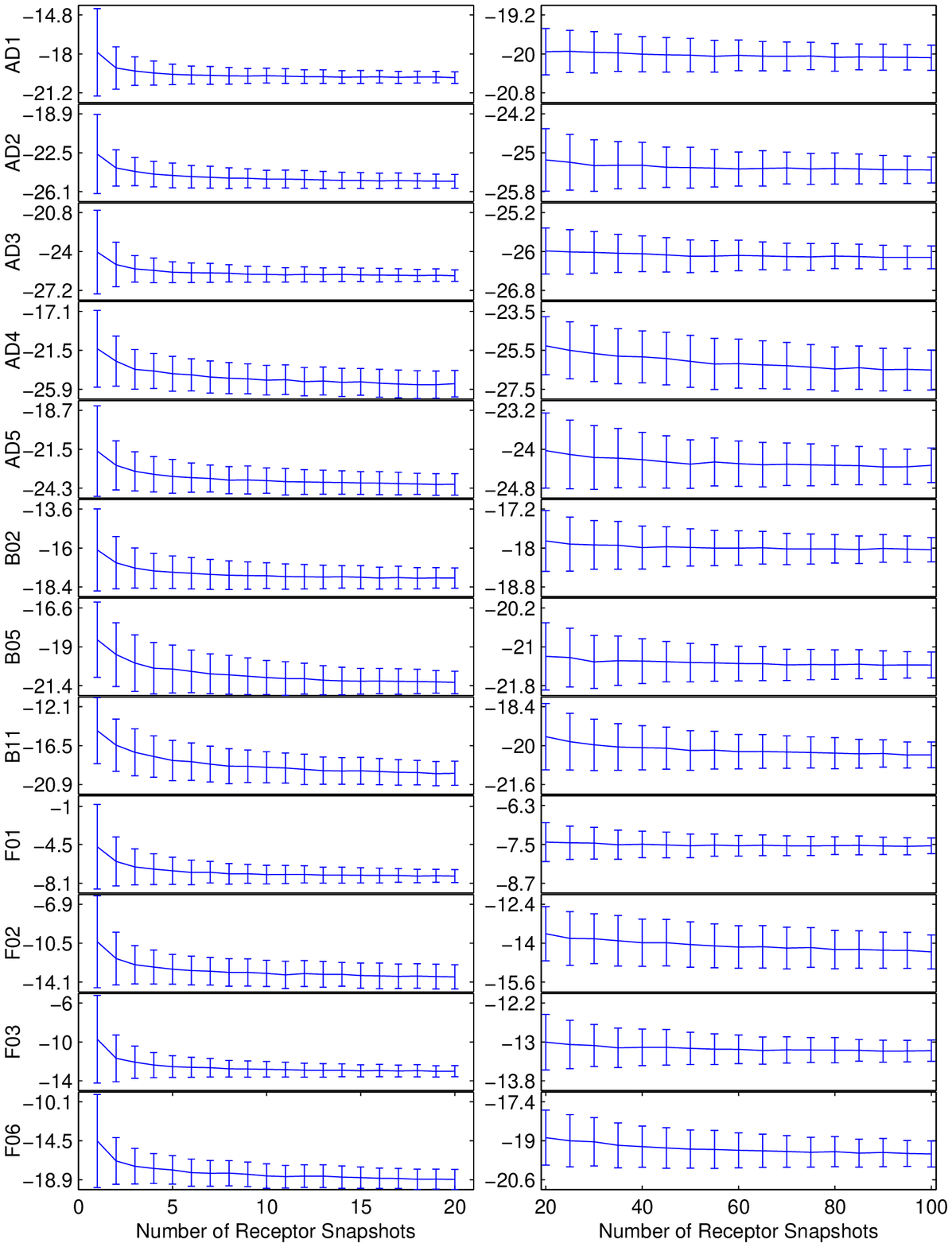}
\caption{
Estimates of the binding free energy $\Delta G^\circ$ of various ligands to Cu[7] (kcal/mol), using PBSA energies, as a function of the number of receptor snapshots.
The line and error bars denote the mean and standard deviation from bootstrapping: the binding free energy is estimated 100 times using random selections of $N$ out of 100 binding PMFs.
\label{fig:S3}}
\end{figure}

\end{document}